%% file: charm_tetraquarks_paper.tex
\documentclass[11pt,a4paper]{article}
\pdfoutput=1
\usepackage{amssymb}
\usepackage{amsmath}
\usepackage{graphicx}
\usepackage{booktabs}
\usepackage{color}
\usepackage{textcomp}
\usepackage{multirow}
\usepackage{bm}
\usepackage{caption}
\usepackage{subcaption}
\usepackage{tikz}
\usepackage{longtable}
\usepackage{colortbl}
\usepackage{rotating}

\usepackage{jheppub}
\usepackage[utf8]{inputenc}

\begin{document}
\input{title}

\maketitle

\input{introduction}

\input{tetraquark}

\input{spectrum}

\input{results}

\input{stability}

\input{interpretations}

\input{conclusions}

\bigskip

\begin{acknowledgments}
We thank our colleagues with the Hadron Spectrum Collaboration. GKCC acknowledges support from the Cambridge European Trust and St John's College, Cambridge. GKCC and CET acknowledge support from the U.K. Science and Technology Facilities Council (STFC)  [grant number ST/L000385/1]. JJD acknowledges support from the U.S. Department of Energy Early Career award contract DE-SC0006765. JJD and RGE acknowledge support from U.S. Department of Energy contract DE-AC05-06OR23177, under which Jefferson Science Associates, LLC, manages and operates Jefferson Lab. 

This work used the DiRAC Data Analytic system at the University of Cambridge, operated by the University of Cambridge High Performance Computing Service on behalf of the STFC DiRAC HPC Facility (www.dirac.ac.uk). This equipment was funded by BIS National E-infrastructure capital grant ST/K001590/1, STFC capital grants ST/H008861/1 and ST/H00887X/1, and STFC DiRAC Operations grant ST/K00333X/1.
This work also used the DiRAC Complexity system, operated by the University of Leicester IT Services, which forms part of the STFC DiRAC HPC Facility (www.dirac.ac.uk). This equipment is funded by BIS National E-Infrastructure capital grant ST/K000373/1 and STFC DiRAC Operations grant ST/K0003259/1. DiRAC is part of the National E-Infrastructure.
The software codes {\tt Chroma}~\cite{Edwards:2004sx} and {\tt QUDA}~\cite{Clark:2009wm,Babich:2010mu} were used to compute the propagators required for this project on the Wilkes GPU cluster at the University of Cambridge High Performance Computing Service (http://www.hpc.cam.ac.uk/), provided by Dell Inc., NVIDIA and Mellanox, and part funded by STFC with industrial sponsorship from Rolls Royce and Mitsubishi Heavy Industries. Propagators were also computed on clusters at Jefferson Laboratory under the USQCD Initiative and the LQCD ARRA project.
Gauge configurations were generated using resources awarded from the U.S. Department of Energy INCITE program at Oak Ridge National Lab, the NSF Teragrid at the Texas Advanced Computer Center and the Pittsburgh Supercomputer Center, as well as at Jefferson Lab.
\end{acknowledgments}

\appendix
\input{appendix}

\clearpage
\bibliography{charm_tetraquarks_paper}
\bibliographystyle{JHEP}

\end{document}

%% file: title.tex
\title{Tetraquark operators in lattice QCD and exotic flavour states in the charm sector}

\author[a]{Gavin~K.~C.~Cheung,} \emailAdd{gkcc2@damtp.cam.ac.uk}
\author[a]{Christopher~E.~Thomas,} \emailAdd{c.e.thomas@damtp.cam.ac.uk}
\author[b,c]{Jozef~J.~Dudek,} \emailAdd{dudek@jlab.org}
\author[b]{Robert~G.~Edwards} \emailAdd{edwards@jlab.org}
\author{\\(For the Hadron Spectrum Collaboration)}

\affiliation[a]{DAMTP, University of Cambridge, Centre for Mathematical Sciences, Wilberforce Road, Cambridge CB3 0WA, UK}
\affiliation[b]{Thomas Jefferson National Accelerator Facility, 12000 Jefferson Avenue, Newport News, VA 23606, USA}
\affiliation[c]{Department of Physics, College of William and Mary, 300 Ukrop Way, Williamsburg, VA 23187, USA}

\abstract{We present a general class of operators resembling compact tetraquarks which have a range of colour-flavour-spin structures, transform irreducibly under the symmetries of the lattice and respect other relevant symmetries. These constructions are demonstrated in lattice QCD calculations with light quarks corresponding to $m_\pi = 391$ MeV. Using the distillation framework, correlation functions involving large bases of meson-meson and tetraquark operators are computed in the isospin-1 hidden-charm and doubly-charmed sectors, and finite-volume spectra are extracted with the variational method. We find the spectra are insensitive to the addition of tetraquark operators to the bases of meson-meson operators.
For the first time, through using diverse bases of meson-meson operators, the multiple energy levels associated with meson-meson levels which would be degenerate in the non-interacting limit are extracted reliably.
The number of energy levels in each spectrum is found to be equal to the number of expected non-interacting meson-meson levels in the energy region considered and the majority of energies lie close to the non-interacting levels. Therefore, there is no strong indication for any bound state or narrow resonance in the channels we study. }

\preprint{\begin{tabular}{r}DAMTP-2017-33\\JLAB-THY-17-2541\end{tabular}}
\arxivnumber{1709.01417}

%% file: introduction.tex
\section{Introduction}\label{sec:introduction}

There are an abundance of experimentally-observed mesons containing one or more heavy quarks~\cite{Olive:2016xmw} and these provide a window on a rich variety of strong-interaction physics. In particular, many of them, the so-called `$X,Y,Z$'s', are not compatible with quark model expectations. Clear examples of this incompatibility are the charged charmonium-like $Z_c^+(3900)$ and $Z_c^+(4430)$ which cannot be solely $c\bar{c}$ and must contain at least one additional quark-antiquark pair. One possible explanation for such exotic states is that they are tetraquarks, compact bound states of four quarks. Others, for example Ref.~\cite{Rupp:2016jdk}, suggest that compact tetraquarks are not required to explain the observed spectrum. Recent reviews of some of the $X,Y,Z$'s, with interpretations such as compact tetraquarks, molecular mesons, hybrid mesons and threshold cusps, can be found in Ref.~\cite{Brambilla:2014jmp,Chen:2016qju,Esposito:2016noz,Lebed:2016hpi,Guo:2017jvc, Olsen:2017bmm}. As well as hidden-charm $c\bar{c}q\bar{q}$ configurations, doubly-charmed $cc\bar{q}\bar{q}$ tetraquarks have been hypothesised~\cite{Moinester:1995fk, PhysRevD.71.014008, Hyodo:2012pm, Esposito:2013fma}, but there are currently no experimental candidates for these. 

Quantum Chromodynamics (QCD) is the fundamental theory of the strong interaction and, in principle, should predict whether four-quark states exist and whether these are consistent with the expectations of tetraquark models or other interpretations. The only ab-initio framework for performing systematically-improvable calculations at the hadronic scale is lattice QCD: spacetime is discretised on a finite four-dimensional Euclidean lattice and Monte Carlo techniques are used to compute correlation functions from which observables can be extracted. The discrete spectrum of finite-volume energy eigenstates is obtained from calculations of two-point correlation functions involving interpolating operators which have the required quantum numbers. Hadron-hadron scattering amplitudes, and hence the properties of resonances and other scattering phenomena, can be calculated via the L\"uscher formalism~\cite{Luscher:1990ux,Luscher:1991cf} which relates finite-volume spectra to infinite-volume scattering amplitudes. There is currently no extension of this formalism to three or more hadron scattering channels that is practical to use in calculations but this is an active area where progress is being made. A more in-depth review of the L\"uscher formalism and a discussion on its applications and extensions can be found in Ref.~\cite{Briceno:2017max}.

The Hadron Spectrum Collaboration has developed a range of interpolating operators resembling quark-antiquark~\cite{Dudek:2010,Thomas:2012} and meson-meson~\cite{Dudek:2012gj,Dudek:2012xn} structures which transform irreducibly under the symmetries of the lattice and efficiently interpolate the states of interest. These operators have proven very successful in recent computations of finite-volume spectra which are then used to determine scattering amplitudes~\cite{Dudek:2012gj, Dudek:2012xn,Dudek:2014qha, Wilson:2014cna, Wilson:2015dqa, Dudek:2016cru, Briceno:2016mjc, Moir:2016srx, Briceno:2017qmb}. As has been emphasised in studies such as those in Refs.~\cite{Dudek:2012xn,Wilson:2015dqa}, not including a sufficiently diverse set of relevant operators in the calculations could lead to an unreliable determination of finite-volume spectra and, in turn, incorrect scattering amplitudes. Hence, it is desirable to consider operators with other potentially-relevant colour-flavour-spatial-spin structures, resembling compact tetraquarks, and investigate whether their inclusion has any impact on the extracted spectra. The main goal of this work is to develop a very general class of operators with compact tetraquark structures, which transform irreducibly under the symmetries of the lattice and which respect other relevant symmetries. We will test these constructions in lattice QCD computations of spectra in hidden-charm and doubly-charmed channels. These include isospin-1 $J^{PG} = 0^{+-}, 1^{++}, 1^{+-}$ hidden-charm spectra\footnote{It is important to emphasise that $G$ refers to $G$-parity since $C$-parity is not a good quantum number for charged states. Note that $C = -G$ in isospin-1 for the neutral component.} which are relevant for exotic charged charmonium-like states and where the lightest tetraquark multiplet is expected to appear~\cite{Maiani:2004vq}, and isospin-0 $J^P = 0^+, 1^+, 2^+$ and isospin-$\tfrac{1}{2}$ strange $J^P = 0^+, 1^+$ exotic doubly-charmed spectra.

There have been a number of recent lattice QCD studies of tetraquarks containing one or more heavy quarks. Computations have not found any clear indication for the presence of hidden or open-charm tetraquarks~\cite{Ikeda:2013vwa,Prelovsek:2014swa,Guerrieri:2014nxa,Padmanath:2015era}. Other recent lattice QCD calculations relevant for the channels we study can be found in Refs.~\cite{Chen:2014afa,Chen:2015jwa,Ikeda:2016zwx}. In the bottom sector, there is some evidence supporting the existence of a doubly-bottom $(I)J^P = (0)1^+$ tetraquark where finite-volume spectrum calculations find an energy level below the relevant meson-meson thresholds~\cite{Francis:2016hui}. In addition, a number of computations of the potential between two static quarks in the presence of two light quarks~\cite{Bicudo:2012qt, Brown:2012tm, Bicudo:2015vta, Bicudo:2015kna, Peters:2016isf, Bicudo:2016ooe, Bicudo:2017szl} have found evidence for a bound state~\cite{Bicudo:2012qt,Bicudo:2015vta, Peters:2016isf, Bicudo:2016ooe, Bicudo:2017szl}. We discuss these studies further in the context of our results in Section~\ref{sec:interpretations}.

The structure of the rest of this paper is as follows.  We begin in Section~\ref{sec:tetraquark} by describing the construction of a general class of tetraquark operators which transform irreducibly under the symmetries of the lattice. In Section~\ref{sec:methodology}, the methodology for calculating the finite-volume spectrum with large bases of operators in the distillation framework is presented. The resulting spectra in the hidden-charm and doubly-charmed sectors are presented in Section~\ref{sec:results}. Some systematic effects and the stability of the extracted spectra are investigated in Section~\ref{sec:stability}. We discuss the results in light of phenomenological and other lattice QCD studies in Section~\ref{sec:interpretations} before giving a summary in Section~\ref{sec:summary}. Appendices present some additional properties of diquark and tetraquark operators, give quark model interpretations of the diquark structures and list the operators used to calculate the finite-volume spectra.

%% file: tetraquark.tex
\section{Tetraquark operator construction}\label{sec:tetraquark}

To construct interpolating operators which resemble a compact tetraquark, we combine a diquark operator with an anti-diquark operator. The diquark operator is built from two quark fields coupled together to obtain appropriate colour, flavour and spin quantum numbers and, analogously, the anti-diquark operator is built from two antiquarks. The diquark and anti-diquark are then combined to form a colour singlet with the desired flavour and spin. These constructions provide, with no loss in generality, a convenient way to build a diverse class of \emph{tetraquark operators} which have the required quantum numbers and respect appropriate symmetries. In this section we present an overview of these operators: we begin by describing the flavour and colour structures before presenting expressions for the diquark, anti-diquark and tetraquark operators. Further details and a discussion of some additional properties can be found in Appendix~\ref{app:tetraquark}. Further model-dependent understanding on how the different diquark configurations interpolate different states can be found in Appendices~\ref{subsec:onegluon} and \ref{subsec:nonrelativistic}.

The diquark operator is constructed by coupling two quark fields together to definite colour, flavour and continuum spin. In colour space, the quarks belong in the fundamental representation of SU$(3)_C$ and so the diquark is in either the antisymmetric $\underline{\bar{3}}$ representation or the symmetric $\underline{6}$ representation. In flavour space, we use SU$(3)_F$ constructions to form a convenient basis of operators, but this does not imply any assumption of SU$(3)_F$ symmetry in the theory -- as long as a sufficient basis is used, an arbitrary flavour combination can be constructed from a linear combination of these operators. The up, down and strange ($u,d$ and $s$) quarks belong in the fundamental representation of SU$(3)_F$ and the charm ($c$) quark is placed in a singlet. The quarks are coupled together to obtain the desired flavour irrep out of $\underline{1}, \underline{3}, \underline{\bar{3}}$ and $\underline{6}$. For example, coupling two $u,d,s$ quarks as $\underline{3} \otimes \underline{3} \rightarrow \bar{\underline{3}}$, this irrep gives a component with flavour quantum numbers (isospin, strangeness) $= (0,0)$ with flavour structure
$\frac{1}{\sqrt{2}} \left(u d - d u \right)$, and a $ (\frac{1}{2},-1)$ multiplet with flavour structure $ \frac{1}{\sqrt{2}} \left(us - su \right)$ and $\frac{1}{\sqrt{2}} \left(ds - s d \right)$. Alternatively, coupling a $c$ quark with a $u,d,s$ quark as $\underline{1} \otimes \underline{3} \rightarrow \underline{3}$ gives a $(0,-1)$ component with flavour structure $c s$, and a $ (\frac{1}{2}, 0)$ multiplet with flavour structure $c u$ and $c d$. If both quarks are in the same flavour representation, Fermi symmetry requires that the overall operator is antisymmetric under the interchange of the quarks and this constrains the allowed diquark configurations.

The diquark operator in colour irrep $R$ (row $r$), flavour irrep $F$ (row $f$) and continuum spin $J$ ($J_z$ component $m$) is,
\begin{equation}
\delta^{J[\Gamma]}_{ RF; rfm}(\vec{x},t) = \sum_{r_a,r_b} \langle \underline{3}, r_a; \underline{3}, r_b | R, r \rangle \sum_{f_a,f_b} \langle F_a, f_a; F_b, f_b | F, f\rangle \ q_{F_a;r_a f_a}^T(\vec{x},t) C \Gamma_m q_{F_b;r_b f_b} (\vec{x},t)
\label{eqn:diquark}
\end{equation}
where spinor indices have been suppressed, $q(\vec{x},t)$ is a quark field smeared with the distillation operator as discussed in Section~\ref{subsec:distillation}, $\langle D_a, d_a; D_b, d_b | D, d \rangle$ are the SU(2) or SU(3) Clebsch-Gordan coefficients that couple the irreps $D_a \otimes D_a \rightarrow D$ with $d_a$, $d_b$ and $d$ the irrep rows, $C$ is the charge conjugation matrix such that $\gamma_0 = C \gamma_0^T C$, and $\Gamma$ is a Dirac gamma matrix which determines $J$, $m$ and other properties of the diquark operator as shown in Table~\ref{tab:gamma} in Appendix~\ref{app:tetraquark}. Choosing an appropriate $\Gamma$ gives access to spins up to $J=1$ -- in order to access higher spins or excitations, this operator can be generalised by including gauge-covariant derivatives in a similar way to the fermion-bilinear operator constructions discussed in Ref.~\cite{Dudek:2010}.

In the anti-diquark operator, the antiquarks belong in the anti-fundamental representation of SU$(3)_C$ and therefore couple to colour irrep $\underline{3}$ or $\bar{\underline{6}}$. The up, down and strange antiquarks belong in the $\underline{\bar{3}}$ irrep of SU$(3)_F$ and the charm antiquark is in the singlet. Possible flavour irreps for the anti-diquark are therefore $\underline{1}, \underline{3}, \underline{\bar{3}}$ and $\underline{\bar{6}}$. If both antiquarks are in the same flavour irrep, Fermi symmetry again constrains the allowed configurations. The anti-diquark operator is defined in an analogous way to the diquark operator as,
\begin{equation}
\bar{\delta}^{J[\Gamma]}_{RF; rfm}(\vec{x},t) =  \sum_{r_a,r_b} \langle \underline{\bar{3}}, r_a; \underline{\bar{3}}, r_b | R, r \rangle \sum_{f_a,f_b} \langle F_a, f_a; F_b, f_b | F, f\rangle  \ \bar{q}_{F_a;r_af_a}(\vec{x},t) \Gamma_m C \bar{q}_{F_b;r_bf_b}^T(\vec{x},t)
\end{equation}
where $\bar{q}(\vec{x},t)$ is a (smeared) antiquark field. Here the charge conjugation matrix comes after the Dirac gamma matrix so that under the charge conjugation operator ${\mathcal{C} \, \bar{\delta}^{J[\Gamma]}_{RF} \, \mathcal{C}^{-1} = \delta^{J[\Gamma]}_{\bar{R}\bar{F}}}$, and this ensures a convenient definition of tetraquark operators with definite $G$-parity as discussed later. 

Tetraquark operators are formed by coupling a diquark operator and an anti-diquark operator to a colour singlet with definite flavour and spin.  The only possible diquark and anti-diquark colour combinations which give a colour singlet are $\underline{\bar{3}} \otimes \underline{3}$ and $\underline{6} \otimes \underline{\bar{6}}$ and this restricts the possible diquark--anti-diquark configurations. The flavour quantum numbers of the tetraquark operator are obtained by coupling the appropriate flavour irreps of the diquark and anti-diquark and then choosing the desired row. By projecting onto zero momentum, tetraquark operators have definite parity and, in channels where $G$-parity is a good quantum number, operators with definite $G$-parity can be constructed. The tetraquark operator, projected onto momentum $\vec{p}$, is,
\begin{equation}
\begin{aligned}
\mathcal{T}^{J[\Gamma_1,\Gamma_2]}_{[R_1,R_2]F[F_1,F_2];fm}(\vec{p},t) = \sum_{\vec{x}} e^{i\vec{p} \cdot \vec{x}} \ \sum_{m_1,m_2} \langle J_1,m_1;J_2,m_2 | J,m  \rangle \ \sum_{r_1,r_2} \langle R_1, r_1; R_2, r_2 | \underline{1}  \rangle \\ 
\times \sum_{f_1,f_2} \langle F_1, f_1; F_2, f_2 | F,f  \rangle \  \delta^{J_1[\Gamma_1]}_{R_1 F_1; r_1 f_1 m_1} (\vec{x},t) \ \bar{\delta}^{J_2[\Gamma_2]}_{R_2 F_2; r_2 f_2 m_2} (\vec{x},t) \, .
\end{aligned}
\label{eqn:tet}
\end{equation}
For the remainder of this study, we only consider $\vec{p} = 0$ and so this operator has definite parity $P$ which is determined by the gamma matrices $\Gamma_1$ and $\Gamma_2$ as described in Appendix~\ref{app:tetraquark}. In channels where $G$-parity is a good quantum number, a tetraquark operator with definite $G$-parity is given by, 
\begin{equation}
\mathcal{T}^{J[\Gamma_1,\Gamma_2],P G}_{[R_1,R_2]F[F_1,F_2];fm}(\vec{p}=\vec{0},t) = \mathcal{T}^{J[\Gamma_1,\Gamma_2]}_{[R_1,R_2]F[F_1,F_2];fm}(\vec{0},t) + \tilde{G} \; \mathcal{T}^{J[\Gamma_2,\Gamma_1]}_{[\bar{R}_2,\bar{R}_1]F[\bar{F}_2,\bar{F}_1];fm}(\vec{0},t) \, ,
\label{eqn:tetragparity}
\end{equation}
where $\tilde{G} = \pm 1 $.  The $G$-parity of this operator is $G = \tilde{G} \xi_J \xi_1 \xi_3$ where $\xi_J, \xi_1, \xi_3$ are phases arising from the exchange symmetry of the Clebsch-Gordan coefficients in Equation~\eqref{eqn:tet} as described explicitly in Appendix~\ref{app:tetraquark}. 

This tetraquark operator has definite spin $J$ in the continuum but the lattice discretisation breaks rotational symmetry and so $J$ is no longer a good quantum number.  With a cubic lattice discretisation and volume, the symmetry group is reduced to the octahedral group $\mathrm{O}_h$ for states at rest~\cite{Johnson:1982yq} and broken further to the little group for states with non-zero momentum~\cite{Moore:2005dw}.  The distribution, or \emph{subduction}, of $J \leq 4$ into the irreps of $\mathrm{O}_h$ is tabulated in Table~\ref{tab:subduction}.  We construct lattice tetraquark operators which transform in lattice irrep $\Lambda$ (row $\mu$) by subducing the continuum operators as described in Ref.~\cite{Dudek:2010},
 \begin{equation}
\mathcal{T}^{\Lambda[J[\Gamma_1,\Gamma_2]],P (G)}_{[R_1,R_2]F[F_1,F_2];f\mu}(\vec{p}=\vec{0},t) = \sum_m S_{\Lambda, \mu}^{J,m} \ \mathcal{T}^{J[\Gamma_1,\Gamma_2],P (G)}_{[R_1,R_2]F[F_1,F_2];fm}(\vec{p}=\vec{0},t) \, ,
\end{equation}
where $S$ are subduction coefficients. The generalisation to $\vec{p} \neq \vec{0}$ involves the construction of helicity operators and then the subduction to irreps of the little group of $\vec{p}$ as discussed in Ref.~\cite{Thomas:2012}. We will use these lattice tetraquark operators to calculate correlation functions in lattice QCD from which the finite-volume spectrum can be extracted.

\begin{table}
\begin{center}
\begin{tabular}{ c| c}
$J$ & $\Lambda$ \\
\hline
0 & $A_1$ \\
1 & $T_1$ \\
2 & $E \oplus T_2$ \\
3 & $A_2 \oplus T_1 \oplus T_2$ \\
4 & $A_1 \oplus E\oplus T_1 \oplus T_2$\\
\end{tabular}
\caption{Subduction of continuum spin $J$ into lattice irreps $\Lambda$ of $O_h$ for $J\leq 4$.}
\label{tab:subduction}
\end{center}
\end{table}

%% file: spectrum.tex
\section{Calculation of the spectrum}\label{sec:methodology}

To determine the spectrum in each quantum-number channel, we calculate a matrix of two-point correlation functions using a basis of interpolating operators with appropriate quantum numbers,
\begin{equation}
C_{ij}(t) = \langle 0 | \mathcal{O}_{i}^{}(t) \mathcal{O}_{j}^{\dagger}(0) | 0 \rangle \, ,
\label{eqn:corr}
\end{equation}
between a creation operator $\mathcal{O}_j^\dagger(0)$ at the source with Euclidean time 0 and an annihilation operator $\mathcal{O}_i(t)$ at the sink with Euclidean time $t$. Inserting a complete set of energy eigenstates into this expression gives,
\begin{equation}
C_{ij}(t) = \sum_\mathfrak{n} \frac{1}{2E_\mathfrak{n}} Z_i^{\mathfrak{n}\ast} Z_j^\mathfrak{n} e^{-E_\mathfrak{n} t} \, ,
\label{eqn:2ptspectra}
\end{equation}
where $|\mathfrak{n}\rangle$ is an energy eigenstate with energy $E_\mathfrak{n}$ and the operator-state matrix elements, $Z_i^\mathfrak{n} \equiv \langle \mathfrak{n} | \mathcal{O}_i^\dagger(0) | 0 \rangle$, are also referred to as \emph{overlaps}. Note that in a finite volume, the set of energy eigenstates is discrete. The spectrum can be extracted by utilising the variational method \cite{Michael:1985, Luscher:1990, Blossier:2009kd}: a generalised eigenvalue problem $C_{ij}(t)v^\mathfrak{n}_j = \lambda^\mathfrak{n}(t,t_0)C_{ij}(t_0)v^\mathfrak{n}_j$ is solved for some appropriate choice of $t_0$, the eigenvalues $\lambda^\mathfrak{n}$, known as \emph{principal correlators}, are related to $E_{\mathfrak{n}}$ and the eigenvectors $v_i^\mathfrak{n}$ are related to the overlaps. In our implementation of the variational method described in Refs.~\cite{Dudek:2007,Dudek:2010}, we fit the principal correlators to the function,
\begin{equation}
 \lambda^\mathfrak{n}(t,t_0) = (1-A_\mathfrak{n})e^{-E_\mathfrak{n}(t-t_0)} + A_\mathfrak{n}e^{-E_\mathfrak{n}'(t-t_0)} \, ,
\label{eqn:fit}
\end{equation}
where the fit parameters are $E_\mathfrak{n}$, $E_\mathfrak{n}'$, and $A_\mathfrak{n}$. The second exponential is used to account for possible contamination due to excited states. The eigenvectors can be used to construct \emph{optimised operators}, $\Omega_\mathfrak{n}^\dagger \sim  \sum_i v^\mathfrak{n}_i \mathcal{O}_i^\dagger$~\cite{Dudek:2012gj}, the optimal linear combination of the operators that interpolates state $\mathfrak{n}$. As will be discussed later, these optimised operators are useful for the construction of operators resembling pairs of mesons.

In principle, any operator can interpolate every state with the same quantum numbers according to Equation~\eqref{eqn:2ptspectra} and one can use a basis containing the tetraquark operators described above to fully calculate the finite-volume spectrum. However, in practice, previous studies such as Refs.~\cite{Dudek:2012xn,Wilson:2015dqa} have highlighted that a sufficiently diverse set of interpolating operators must be used if finite-volume spectrum is to be extracted reliably. Even in the absence of interactions, the spectra we study will contain meson-meson-like states or admixtures of such states (and other multi-hadron combinations at higher energies). It has been show in previous work~\cite{Dudek:2012gj,Dudek:2012xn,Dudek:2014qha,Wilson:2014cna,Wilson:2015dqa,Moir:2016srx,Dudek:2016cru,Briceno:2016mjc, Briceno:2017qmb} that such states can be efficiently interpolated by including meson-meson-like operators. Therefore, to efficiently and reliably extract the finite-volume spectra, our operator bases will contain operators of meson-meson and tetraquark structure. 

\subsection{Meson-meson operators}
\label{subsec:mesonmeson}
In this section we briefly review how operators with a meson-meson-like structure can be constructed from the product of two single-meson-like operators -- further details are given in Refs.~\cite{Dudek:2012gj,Dudek:2012xn}.

Following Refs.~\cite{Dudek:2010,Thomas:2012}, fermion-bilinear operators of continuum spin $J$ and momentum $\vec{p}$ are constructed as,
\begin{equation}
\mathcal{O}^{J,m}(\vec{p},t) = \sum_{\vec{x}} e^{i\vec{p} \cdot \vec{x}} \  \bar{q}(\vec{x},t) [\Gamma \overleftrightarrow{D} \dots \overleftrightarrow{D}]^{J,m} q(\vec{x},t) \, ,
\end{equation}
where we have suppressed colour, flavour and spinor indices for clarity. The quark and antiquark fields are distillation-smeared as discussed below, the quark and antiquark flavour representations are chosen and coupled to give the desired flavour quantum numbers, and $[\Gamma \overleftrightarrow{D} \dots \overleftrightarrow{D}]^{J,m}$ consists of a Dirac gamma matrix $\Gamma$ and gauge-covariant derivatives $\overleftrightarrow{D}$ coupled together to give spin $J$. In the case when $\vec{p} = 0$, $m$ refers to the $J_z$ component, whilst when $\vec{p} \neq 0$ we construct helicity operators using Wigner-D matrices as described in Ref.~\cite{Thomas:2012} and $m$ then refers to the helicity. Since we are working in a finite cubic spatial volume of extent $L$ with periodic boundary conditions, the momentum is quantised to ${\vec{p} = \frac{2\pi}{L}(n_x,n_y,n_z)}$ where $(n_x,n_y,n_z)$ is a triplet of integers -- we use $[n_x n_y n_z]$ as a shorthand notation to denote $\vec{p}$. As for tetraquark operators, lattice operators which transform in lattice irrep $\Lambda$ (row $\mu$) are constructed by subducing the continuum operators to obtain $\mathcal{O}^{[J]}_{\Lambda, \mu}(\vec{p},t) = \sum_m S^{J,m}_{\Lambda,\mu} \mathcal{O}^{J,m}(\vec{p},t)$. We refer to these as \emph{single-meson operators}. 

\emph{Meson-meson operators}~\cite{Dudek:2012gj,Dudek:2012xn} are built from products of two single-meson operators,
\begin{equation}
\mathcal{O}_{\Lambda,\mu}^P(\vec{p}=\vec{0},t) = \sum_{\mu_1,\mu_2,\hat{q}} \mathbb{C}(\vec{p}=\vec{0},\Lambda^P,\mu; \vec{q} , \Lambda_1 , \mu_1; -\vec{q} , \Lambda_2 , \mu_2)
\; \Omega^{M_1}_{\Lambda_1,\mu_1}(\vec{q},t) \; \Omega^{M_2}_{\Lambda_2,\mu_2}(-\vec{q},t) \, ,
\label{eqn:mesonmeson}
\end{equation}
where we have restricted to overall zero momentum, $\Omega^{M_i}_{\Lambda_i, \mu_i}$ is an optimised operator for interpolating meson $M_i$ transforming in the lattice irrep $\Lambda_i$ (row $\mu_i$) and the sum runs over the lattice irrep rows and all momentum directions $\hat{q}$ related by an allowed lattice rotation to couple $\Lambda_1 (\vec{q}) \otimes \Lambda_2 (-\vec{q}) \to \Lambda^P (\vec{p} = \vec{0})$ using generalised Clebsch-Gordan coefficients $\mathbb{C}$. The construction of these meson-meson operators follows the methodology given in Refs~\cite{Dudek:2012gj,Dudek:2012xn} and a more detailed discussion of operators containing mesons with non-zero spin will be presented in a forthcoming publication. Analogous constructions can be used for meson-meson operators with overall non-zero momentum, but in this study we only calculate spectra at overall zero momentum.

A guide to which meson-meson operators should be included in the basis is given by the non-interacting meson-meson energy levels in the energy regions we consider. These are calculated from the relativistic dispersion relation $E = \sqrt{ m_1^2 + \vec{p}_1^{\,2}} + \sqrt{m_2^2 + \vec{p}_2^{\,2}}$ for stable single-mesons. In cases when a single-meson has non-zero spin, there can be multiple ways to couple the orbital and spin angular momenta together to a given meson-meson $J^P$ which subduce into the same irrep, leading to degenerate levels in the non-interacting limit. For example, a pseudoscalar and vector can be coupled to $J^P = 1^+$ in either $s$-wave or $d$-wave. To see how this manifests on the lattice, consider the pseudoscalar with $\vec{p} = [100]$ and the vector with $\vec{p} = [-100]$ coupled to the lattice irrep $\Lambda^P = T_1^+$. The mesons transform in the irreps of the little group $\text{Dic}_4$: the pseudoscalar subduces into the $A_2$ irrep while the helicity-0 and helicity-1 components of the vector subduce into respectively the $A_1$ and $E_2$ irreps. It is possible to obtain $\Lambda^P = T_1^+$ from both $A_2 \otimes A_1$ and $A_2 \otimes E_2$ \cite{Moore:2006ng} and two different linear combinations of these would correspond to $s$ and $d$ wave in the continuum and infinite volume limit. In general, we must include a sufficient number of relevant meson-meson operators that are capable of extracting and disentangling these multiple energy levels. The comparison of spectra calculated with different operator bases in Section~\ref{sec:stability} demonstrates the importance of including a sufficient basis of meson-meson operators.

\subsection{Calculation of correlation functions}
\label{subsec:distillation}

We choose a basis of meson-meson operators as described above and tetraquark operators as described later in Section~\ref{sec:results} and compute the two-point correlation functions using the \emph{distillation} framework~\cite{Peardon:2009}. The combination of distillation and the techniques described here have been demonstrated in, for example, Refs.~\cite{Dudek:2010,Dudek:2011, Dudek:2012gj, Dudek:2012xn,Liu:2012, Wilson:2014cna, Wilson:2015dqa, Dudek:2016cru, Briceno:2016mjc, Moir:2016srx}. In brief, the distillation operator on timeslice $t$ which acts in 3-space ($\vec{x}$ and $\vec{y}$) and colour space ($r$ and $s$) is defined as, $\Box(\vec{x}r,\vec{y}s;t) = \sum_{n=1}^{N_{\text{vecs}}} \xi_n(\vec{x}r;t) \xi_n^\dagger(\vec{y}s;t)$, where in the implementation used here $\xi_n$ are the lowest $N_{\text{vecs}}$ eigenvectors of the gauge-covariant Laplacian. The quark and antiquark fields in interpolating operators are smeared by the distillation operator, $q \rightarrow \Box q$, which removes high-frequency modes and increases the overlap onto lower-lying states. Distillation also allows for a factorisation of the two-point correlation functions as contractions of \emph{perambulators}, $\tau_{nm}(t',t) = \xi^\dagger_n(t') M^{-1} (t',t) \xi_m(t)$, where $M$ is the Dirac matrix, and \emph{elementals} which describe the operators with various structures projected onto definite momentum, and this enables the efficient computation of the correlation function matrix for a large basis of operators. Meson elementals $\Phi^{\alpha \beta}_{n_1 n_2}(\vec{p},t)$ are presented in Ref.~\cite{Dudek:2012xn} and tetraquark elementals are given by,
\begin{equation}
\Psi^{\alpha \beta \gamma \delta}_{n_1 n_2 n_3 n_4}(\vec{p}=\vec{0}, t) = \sum_{\vec{w}p, \vec{x}q, \vec{y}r, \vec{z}s} \mathbb{C}_{pqrs} \ \xi_{n_3}^\dagger(\vec{w}p;t) \xi_{n_4}^\dagger(\vec{x}q;t) (C \Gamma_1)^{\alpha \beta} (\Gamma_2 C)^{\gamma \delta}  \xi_{n_1} (\vec{y} r;t) \xi_{n_2} (\vec{z} s;t) \, ,
\end{equation} 
where $n_i$ index distillation vectors, Greek letters label the Dirac spinor indices and $\mathbb{C}_{pqrs}$ are combinations of SU$(3)_C$ Clebsch-Gordan coefficients that couple the colour representations $\underline{\bar{3}} \otimes \underline{\bar{3}} \otimes \underline{3} \otimes \underline{3} \rightarrow \underline{1}$ as in the tetraquark operators in Section~\ref{sec:tetraquark}. Meson elementals are matrices with $(4 N_{\text{vecs}})^2$ independent components and tetraquark elementals are rank-4 with $(4 N_{\text{vecs}})^4$ independent components. This means that the cost of calculations involving tetraquark operators (multiplying and tracing perambulators and elementals) increases rapidly when the number of vectors is increased. Therefore, if the calculation is to be feasible, the number of vectors must not be too large.

To keep the cost of contractions reasonable by using a relatively small number of vectors for tetraquark operators, whilst maintaining a larger number of vectors for other operators, we introduce a second distillation operator, $\tilde{\Box}(\vec{x}r,\vec{y}s;t) = \sum_{\tilde{n}=1}^{\tilde{N}_{\text{vecs}}} \xi_{\tilde{n}}(\vec{x}r;t) \xi_{\tilde{n}}^\dagger(\vec{y}s;t)$, composed of the lowest $\tilde{N}_{\text{vecs}}$ vectors where $\tilde{N}_{\text{vecs}} < N_{\text{vecs}}$. Quark/antiquark fields in tetraquark operators are smeared with $\tilde{\Box}$ whereas those in other operators are smeared with $\Box$. As an example, consider a meson-meson operator given by $\mathcal{O} \sim (\bar{c} \Box \Gamma \Box u) (\bar{d} \Box \Gamma \Box c)$ and a tetraquark operator, $\mathcal{T} \sim \left( (\tilde{\Box} c)^T C\Gamma (\tilde{\Box} u) \right) \left((\bar{c} \tilde{\Box}) \Gamma C (\bar{d} \tilde{\Box})^T\right)$, where we are suppressing various indices and factors which are not relevant for this discussion. One of the connected contributions to the correlation function between these two operators is, schematically,
\begin{equation}
\langle \mathcal{O}(t) \mathcal{T}(0)^\dagger \rangle  \sim \Phi_{n_1 n_2}(t) \Phi_{n_3 n_4}(t) \tau_{\tilde{n}_3 n_1}(0,t) \tau_{\tilde{n}_4 n_3} (0,t) \tau_{n_4 \tilde{n}_1} (t,0) \tau_{n_2 \tilde{n}_2}(t,0) \Psi_{\tilde{n}_1 \tilde{n}_2 \tilde{n}_3 \tilde{n}_4}(0) \, ,
\label{eqn:distillation}
\end{equation} 
where $n_i = 1, \ldots , N_\text{vecs}$, $\tilde{n}_i = 1, \ldots , \tilde{N}_\text{vecs}$, and we have suppressed spinor indices. Here, the perambulators $\tau(t,0)$ are $4 N_{\text{vecs}} \times 4\tilde{N}_{\text{vecs}}$ rectangular matrices, $\Phi(t)$ are $4N_{\text{vecs}} \times 4N_{\text{vecs}}$ square matrices and $\Psi(0)$ is of rank-4 with $(4\tilde{N}_{\text{vecs}})^4$ components. The viability of having a lower number of distillation vectors for tetraquark operators and some tests of varying the number of vectors are discussed in Section~\ref{subsec:varydistillation}. Although not utilised in this study, another possible use of employing more than one distillation operator is to increase the number of operators in the variational basis through including operators with different smearings. 

\begin{figure}
\begin{center}
\includegraphics[width=\textwidth]{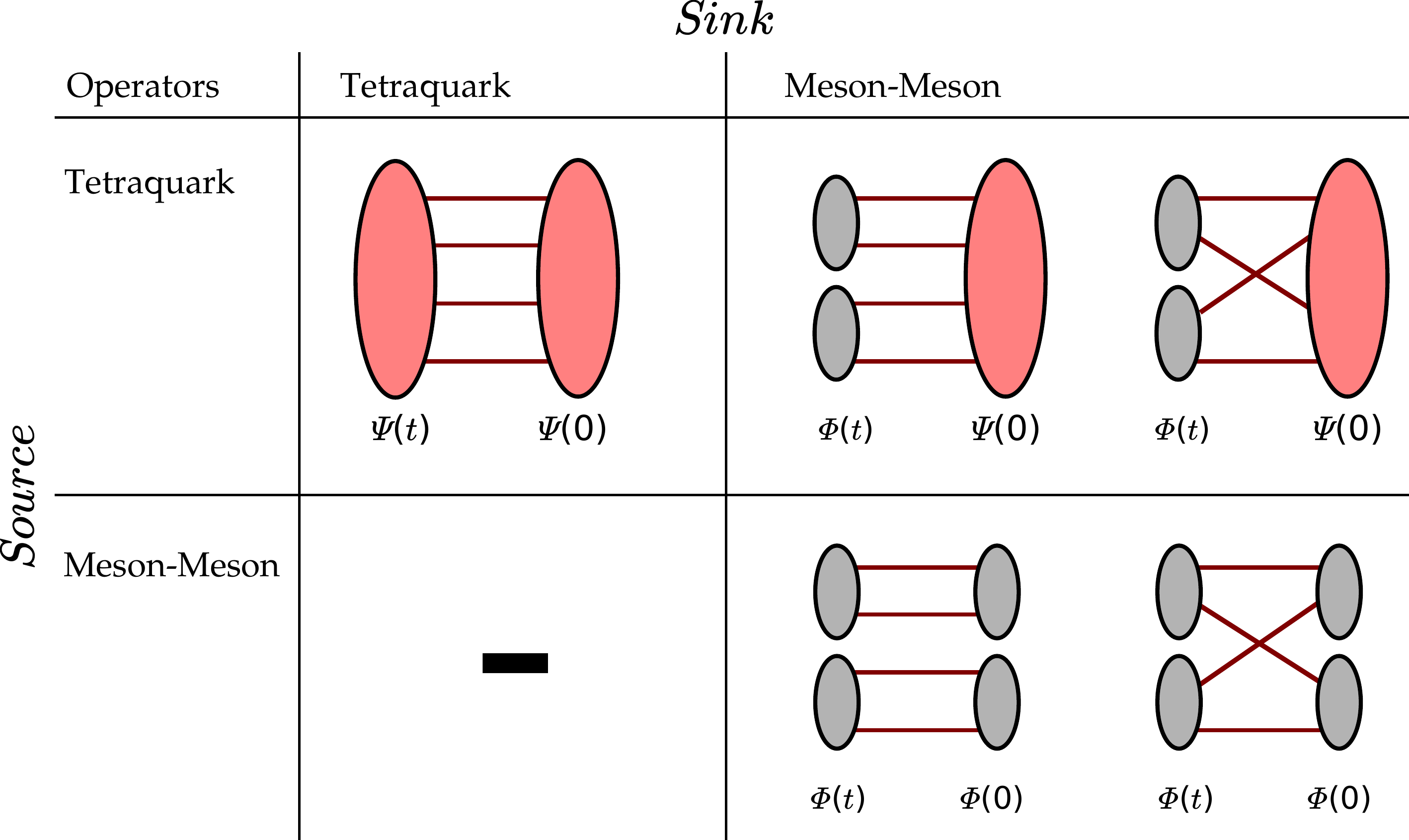}
\caption{A schematic representation of the types of Wick contractions required to compute the two-point correlation function matrices in this study. We use $\Phi$ (grey) to depict the single-meson elementals, $\Psi$ (red) to depict the tetraquark elementals and the lines joining them to depict perambulators.} 
\label{fig:wick}
\end{center}
\end{figure}

To further reduce the computation time, we only calculate one half of the off-diagonal elements in the matrix of two-point correlation functions between a tetraquark operator and meson-meson operator, and then obtain the other half using the Hermiticity of the correlation matrix. In addition, we neglect contributions where a charm quark and antiquark annihilate: these are expected to be small due to OZI suppression and this has been found to be the case empirically in lattice calculations~\cite{Levkova:2010ft}. The elements of the two-point correlation function matrix that we compute are shown in Figure~\ref{fig:wick} where we show a schematic representation of the types of Wick contractions required.

%% file: results.tex
\section{Results} \label{sec:results}

As a first application of these tetraquark operator constructions, we perform calculations on an anisotropic lattice of volume $(L/a_s )^3 \times (T/a_t) = 16^3 \times 128$ where $L$ is the spatial extent of the lattice, $T$ is the temporal extent, $a_s \approx 0.12$ fm is the spatial lattice spacing and $a_t$ is the temporal lattice spacing such that the anisotropy $\xi = \frac{a_s}{a_t} \approx 3.5$. We use 478 configurations generated from a tree level Symanzik improved gauge action and a Clover fermion action with $N_f = 2+1$ flavours of dynamical quarks. The mass parameter of the two degenerate light quarks is such that $m_\pi = 391$ MeV while the strange quark is tuned so that its mass approximates the physical value~\cite{Edwards:2008,Lin:2009}. The quenched Clover charm quark mass parameter is tuned to reproduce the physical $\eta_c$ meson mass \cite{Liu:2012}. When quoting results in physical units, we set the scale using the mass of the $\Omega$ baryon from the measured value on this lattice, $a_t m^{\text{latt.}}_{\Omega} = 0.2951(22)$ \cite{Edwards:2011}, and the experimental mass $m_{\Omega}^{\text{exp.}} = 1672.45(29)$ MeV \cite{Olive:2016xmw}, giving $a_t^{-1} = \frac{m_{\Omega}^{\text{exp.}}}{a_t m_{\Omega}^{\text{latt.}}} = 5667$ MeV. Using several lattice volumes, the anisotropy was measured to be $\xi_{\pi} = 3.444(6)$ from the dispersion relation of the pion \cite{Dudek:2012gj} and $\xi_{D} = 3.454(6)$ from the $D$ \cite{Moir:2013ub}. Using only the $16^3$ volume, we find $\xi_{\eta_c} = 3.484(2)$ from the $\eta_c$. For the purposes of this study, where the anisotropy is only used to compute the location of non-interacting meson-meson energy levels, we will use the value of $\xi_{\pi}$.

To indicate the location of non-interacting meson-meson energy levels on plots, for stable mesons we use the relativistic dispersion relation giving, $E = \sqrt{m_1^2 + \vec{p}_1^{\,2}} + \sqrt{m_2^2 + \vec{p}_2^{\,2}}$, as discussed in Section~\ref{subsec:mesonmeson}. The masses of relevant stable mesons on this lattice ensemble are given in Table~\ref{tab:mesons} and the variationally-optimised operators for these mesons, $\Omega^M_{\Lambda, \mu}$, are constructed from linear combinations of single-meson operators as discussed above. For the $\rho$ meson, which is unstable on this lattice, we compute `non-interacting $M$-$\rho$ energy levels', where $M$ is a stable meson, using the relativistic dispersion relation for $M$ and the finite-volume $\rho$ energy levels obtained on this ensemble as given in Table~\ref{tab:mesons}, i.e.~${E = \sqrt{m_M^2 + \vec{p}_M^{\,2}} + \rho_\Lambda^{\vec{p}}}$. The optimised $\rho$ operators, $\Omega^\rho_{\Lambda, \mu}$, are linear combinations of meson-meson and single-meson operators~\cite{Dudek:2012xn}. It should be emphasised that in this study the only uses of these non-interacting energy levels are to show their location on plots and as an indication for which meson-meson operators should be included in the operator basis.

The meson-meson and tetraquark operators used to calculate the spectrum in each channel are listed in Appendix~\ref{app:ops}. For this lattice volume, we use $\tilde{N}_{\text{vecs}} = 24$ for tetraquark operators and $N_{\text{vecs}} = 64$ for other operators unless stated otherwise. The choice of meson-meson operators has already been discussed in Section~\ref{subsec:mesonmeson}. For the tetraquark operators, ideally all relevant operators of the form described in Section~\ref{sec:tetraquark} would be included in the operator basis, but the computational cost can then become too high for the calculations to be practical for the purposes of this study. In the doubly-charmed isospin-0 $\Lambda^P = A_1^+, E^+, T_2^+$ channels we are able to include all the tetraquark operator constructions allowed by Fermi symmetry. For the remaining channels, we use a subset of tetraquark operators that are expected to overlap onto lower-lying states. Appendices~\ref{subsec:onegluon} and \ref{subsec:nonrelativistic} describe how a non-relativistic model can provide a guide to which diquark/anti-diquark configurations are expected to overlap most efficiently onto lower-lying tetraquarks. In short, tetraquark operators containing a diquark (anti-diquark) with a $\gamma^5$ or $\gamma^0 \gamma^5$ gamma matrix structure and in colour irrep $\underline{\bar{3}}$ $(\underline{3})$ are expected to overlap most efficiently onto a ground-state tetraquark,\footnote{Using Jaffe's terminology, this configuration is commonly known as the `good' diquark.} and at the very least our basis should include such operators. However, it was found that the $\gamma^5$ and $\gamma^0 \gamma^5$ structures do not overlap onto the energy eigenstates in sufficiently distinct ways (the correlation matrix contains approximately linearly-dependent rows/columns). Therefore, instead of having redundant operators, we included a selection of other tetraquark operators to give more diverse bases.

\begin{table}
\begin{center}
\begin{tabular}{c|c}
Meson & Mass (MeV)\\
\hline
$\pi$ & 391.4(7) \\
$D$ & 1885.1(4) \\
$D^\ast$ & 2008.9(6) \\
$D_s$ & 1950.9(3) \\
$D_s^\ast$ & 2071.2(5) \\
$\eta_c$ & 2964.4(2) \\
$J/\psi$ & 3044.7(2) \\
$\chi_{c0}$ & 3426.3(6) \\
\end{tabular}
\hspace{1cm}
\begin{tabular}{c|c}
 & Energy (MeV) \\
\hline
$\rho^{[000]}_{T_1}$ & 890(5)\\
$\rho^{[100]}_{A_1}$ & 1027(4)\\
$\rho^{[100]}_{E_2}$ & 1089(5)\\
 &\\
& \\
& \\
& \\
& \\
\end{tabular}
\caption{Ground state masses of stable mesons (left) and the energy of the lowest-lying finite-volume energy level for lattice irrep $\Lambda$ and momentum relevant for the $\rho$ meson denoted by $\rho_\Lambda^{\vec{p}}$ (right) as measured on our ensemble \cite{Dudek:2012gj, Dudek:2012xn, Liu:2012, Moir:2013ub}. Only the statistical uncertainty is quoted. }
\label{tab:mesons}
\end{center}
\end{table}

We now present computed spectra for a range of channels, beginning with a detailed discussion of the $\Lambda^{PG} = T_1^{++}$ irrep in the isospin-1 hidden-charm sector before presenting other isospin-1 hidden-charm results and then moving to the doubly-charmed sector.

\subsection{Isospin-1 hidden-charm sector}
\label{subsec:hiddencharm}

\begin{figure}[h]
\begin{center}
\includegraphics[width=0.8\textwidth]{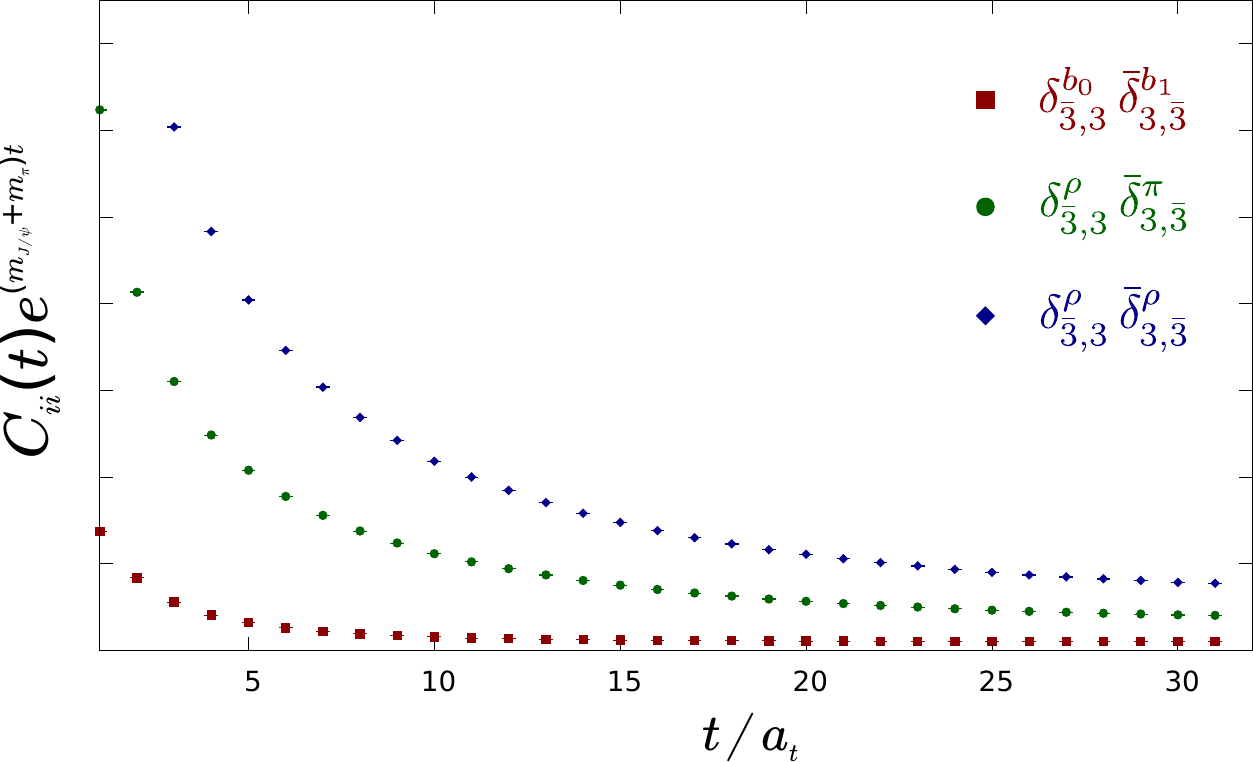}
\caption{$C_{ii}(t) e^{(m_{J/\psi} + m_{\pi})t}$ in arbitrary units for the tetraquark operators given in the legend in the $\Lambda^{PG} = T_1^{++}$ isospin-1 hidden-charm channel. Error bars are smaller than the size of the points shown. }
\label{fig:correlator}
\end{center}
\end{figure}

\begin{figure}[h]
\begin{center}
\includegraphics[width=0.85\textwidth]{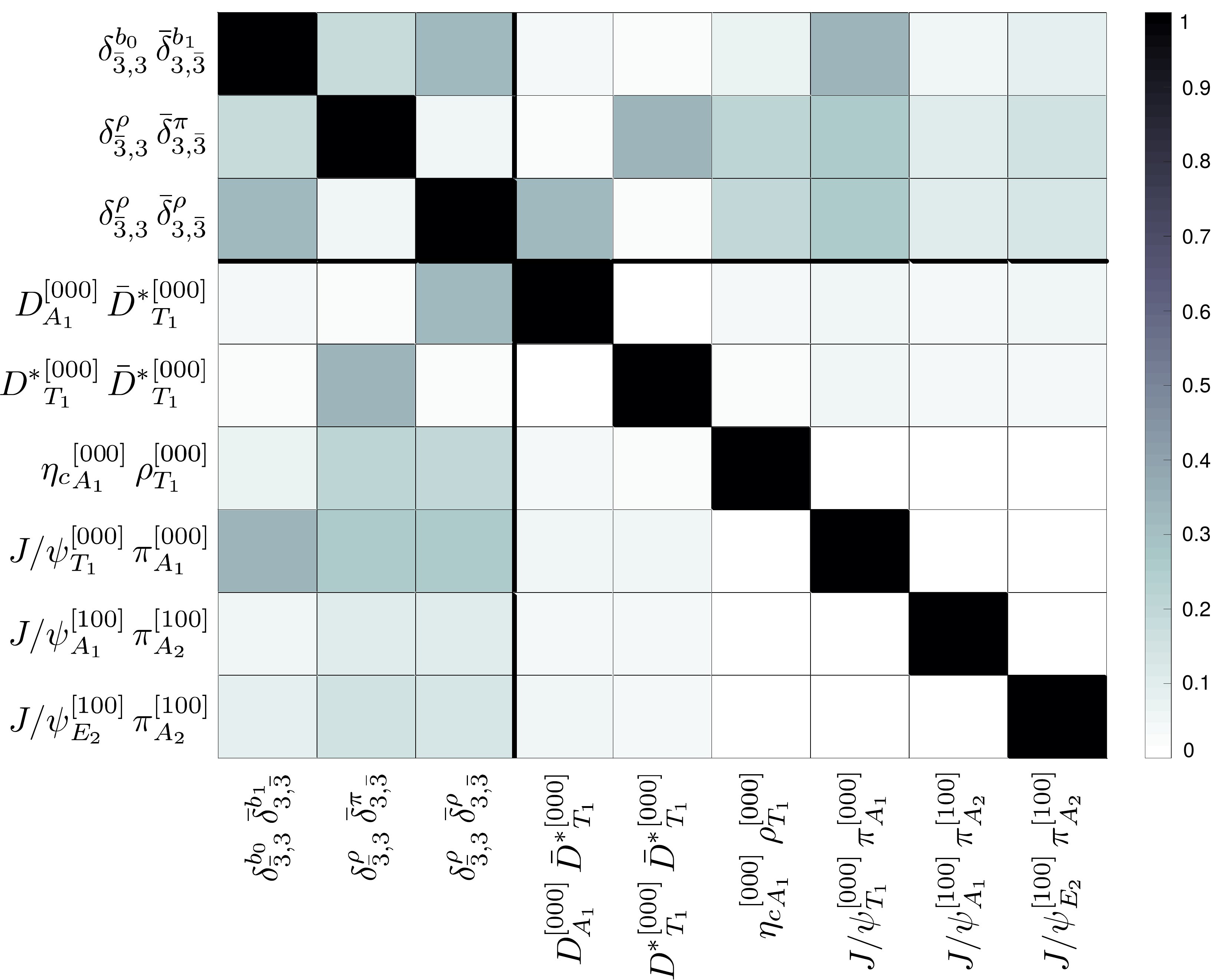}

\caption{Normalised magnitude of elements in the matrix of two-point correlation functions, $|C_{ij}|/\sqrt{C_{ii}C_{jj}}$, on timeslice 3 in the $\Lambda^{PG} = T_1^{++}$ isospin-1 hidden-charm channel. The first three operators are tetraquark operators and the remaining are meson-meson operators ordered as in Table~\ref{tab:hiddencharmops}.} 
\label{fig:heatmap}
\end{center}
\end{figure}

\begin{figure}[!h]
\begin{center}
\includegraphics[width=0.86\textwidth]{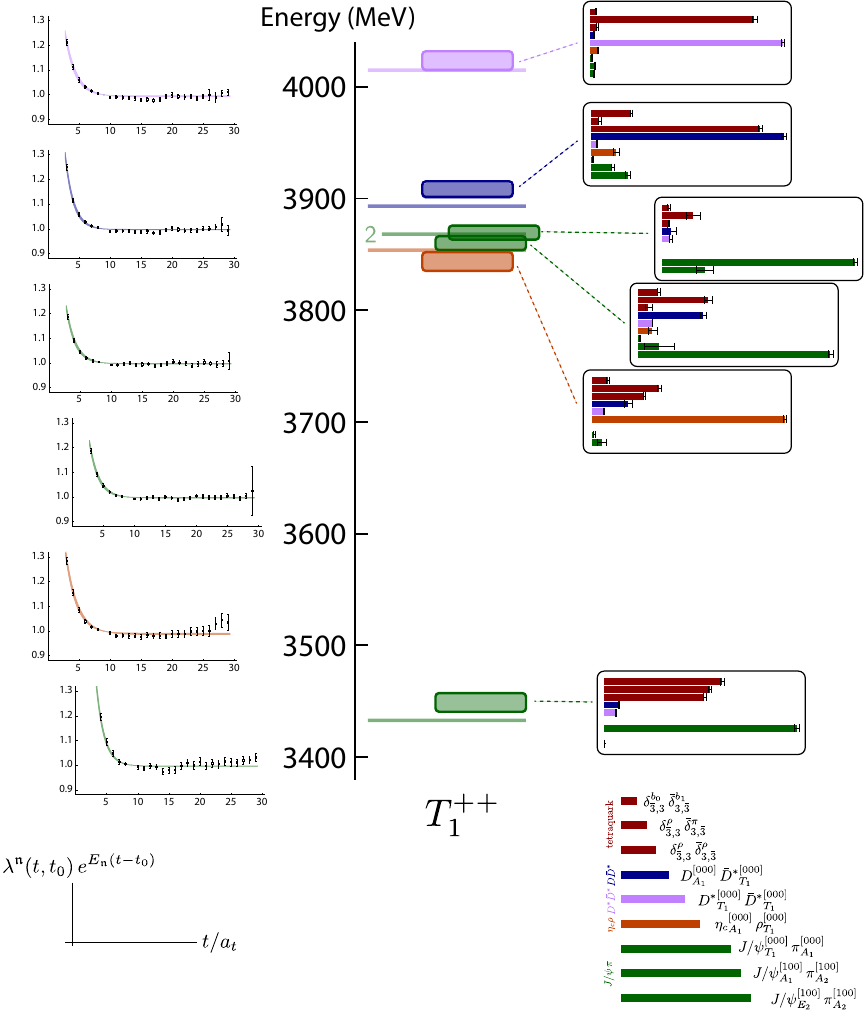}
\caption{The central plot shows the spectrum in the hidden-charm isospin-1 $\Lambda^{PG} = T_1^{++}$ channel calculated using the basis of meson-meson and tetraquark operators given in Table~\ref{tab:hiddencharmops} of Appendix~\ref{app:ops}. Boxes give the computed energies with their vertical extent representing the one-sigma statistical uncertainty on either side of the mean and, solely as a visual aid, they are coloured according to their dominant meson-meson operator overlap. Horizontal lines denote the non-interacting meson-meson energy levels with an adjacent number indicating the degeneracy if it is larger than one. 
The corresponding principal correlators are shown on the left ordered by increasing energy from bottom to top: the data (points) and fits (curves) for $t_0 = 9$ are plotted as $\lambda^{\mathfrak{n}}(t,t_0) e^{E_\mathfrak{n}(t-t_0)}$ showing the central values and one sigma statistical uncertainties; in each case the fit is reasonable with $\chi^2/N_\mathrm{d.o.f} \sim 1$.
The histograms on the right show the operator-state overlaps, $Z_i^\mathfrak{n} = \langle \mathfrak{n} | \mathcal{O}_i^\dagger | 0 \rangle$, for each energy level. The operators are given in the legend and the overlaps are normalised so that the largest value for one given operator across all energy levels is equal to one. }
\label{fig:combined}
\end{center}
\end{figure}

As an illustration of the results, we first discuss some features of the spectrum computed in the $\Lambda^{PG} = T_1^{++}$ irrep\footnote{The lowest spin in this irrep is $J^{PG} = 1^{++}$ and note that $C=-G$ in isospin-1 for the neutral component.} of the isospin-1 hidden-charm sector (with flavour content $c\bar{c} l \bar{l}$ where the light quark and antiquark are coupled to isospin-1). The basis of operators used is given in Table~\ref{tab:hiddencharmops} of Appendix~\ref{app:ops}. Note that, because we do not include contributions arising from a charm quark and antiquark annihilating, our operator bases do not contain any single-meson operators.

The diagonal elements of the matrix of correlators for the three tetraquark operators are shown in Figure~\ref{fig:correlator} -- signals are seen to be precise and significantly non-zero. In Figure~\ref{fig:heatmap} we present the two-point correlator matrix on timeslice $3$. This shows that some of the off-diagonal elements between tetraquark and meson-meson operators are non-zero.

After applying the variational method, principal correlators for the lowest six energy levels are shown in Figure~\ref{fig:combined}. We fit these to Equation~\eqref{eqn:fit} and in each case find a reasonable description having $\chi^2/N_\mathrm{d.o.f} \sim 1$ -- the resulting spectrum is given in the figure. It can be seen that the number of energy levels in the computed spectrum is equal to the number of non-interacting meson-meson levels expected in the energy region considered and they all lie close to the non-interacting levels. As discussed in Section~\ref{subsec:mesonmeson} and indicated in the figure, some of the non-interacting meson-meson energy levels are degenerate. Because our basis of operators has sufficiently different structures, we are able to cleanly extract nearly-degenerate energy levels.

Normalised operator-state overlaps are also shown in Figure~\ref{fig:combined} and we see that every energy level has a dominant overlap onto one meson-meson operator. Additionally, the third and fourth levels have dominant overlaps onto two linearly independent $J/\psi\pi$ operators -- this is not surprising since around this energy there are two degenerate non-interacting levels. We cannot draw strong quantitative conclusions about the tetraquark operator overlaps because the absolute normalisations are somewhat arbitrary and renormalisation factors would be needed to relate the overlaps to physical quantities, but we do see that most states have some overlap onto one or more tetraquark operators.

\begin{figure}[h]
\begin{center}
\includegraphics[width=\textwidth]{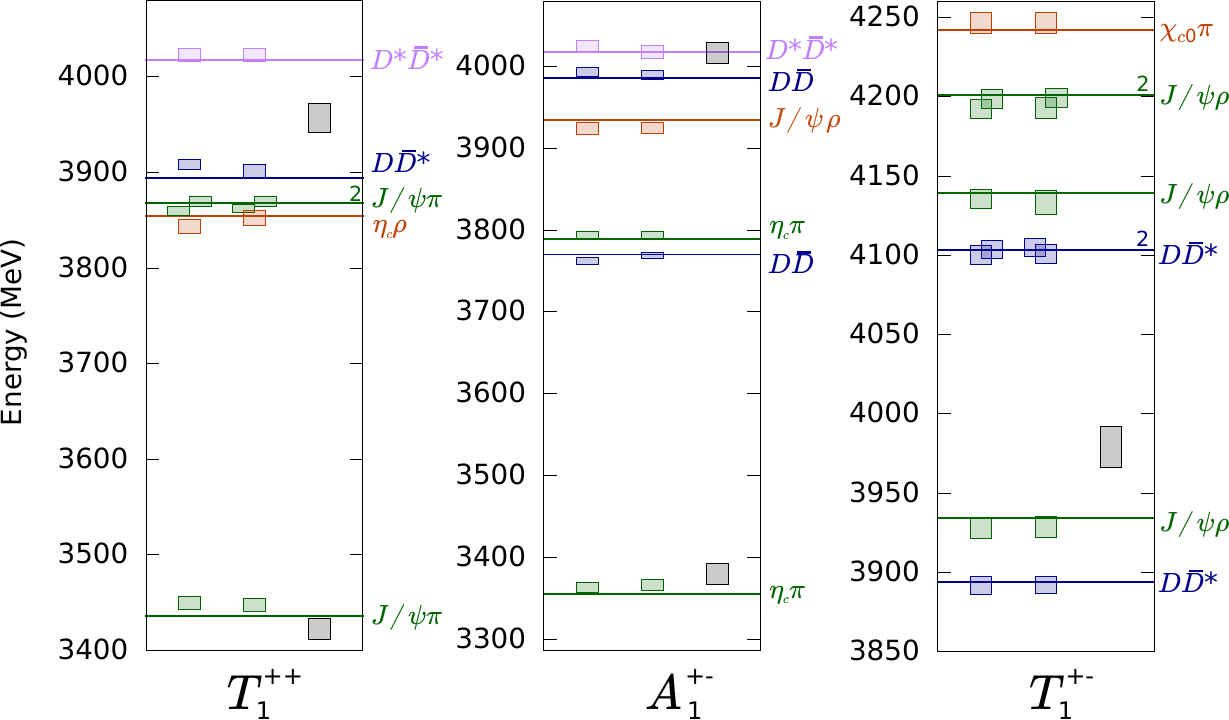}
\caption{As in the spectrum plot of Figure~\ref{fig:combined} but showing the spectra for the isospin-1 hidden-charm sector with $\Lambda^{PG} = T_1^{++}, A_1^{+-},T_1^{+-}$. Within each plot, the left, middle and right column shows the spectrum determined using the full basis of meson-meson and tetraquark operators, only meson-meson operators and only tetraquark operators respectively.}
\label{fig:hiddencharm}
\end{center}
\end{figure}

For comparison, Figure~\ref{fig:hiddencharm} (left panel) shows the $\Lambda^{PG} = T_1^{++}$ spectrum calculated with the full basis of meson-meson and tetraquark operators, with \emph{only meson-meson operators} and with \emph{only tetraquark operators}. No significant deviations are observed between the spectrum computed using the full basis and that computed using the basis of only meson-meson operators. If only tetraquark operators are used, some poorly determined energy levels are found but the spectrum is not reliably extracted and this suggests that these tetraquark operators alone do not constitute a sufficient basis of operators.

Moving to other channels in the isospin-1 hidden-charm sector, extracted spectra for the $\Lambda^{PG} = A_1^{+-},T_1^{+-}$ irreps\footnote{The lowest spin in each of these irreps is respectively $J^{PG} = 0^{+-}, 1^{+-}$.} are shown in Figure~\ref{fig:hiddencharm}. In general, a similar pattern of features is seen as was found for $\Lambda^{PG} = T_1^{++}$: there are no significant deviations between the spectra calculated using the full basis and using only meson-meson operators, the spectrum is not reliably determined if only tetraquark operators are used, and with a full basis of operators the number of energy levels is equal to the number of non-interacting meson-meson levels expected and they lie close to the non-interacting levels.  Furthermore, the operator-state overlaps follow the same qualitative pattern as shown in Figure~\ref{fig:combined}.

From previous studies, when a narrow resonance is present in elastic scattering, an `extra' finite-volume energy level is observed in that energy region but no evidence for such an extra level is seen in our spectra. The results suggest that there are only weak hadron-hadron interactions and no strong indications of a bound state or narrow resonance in these channels. However, the situation is not as straightforward when one considers coupled-channel scattering or broad resonances~\cite{Dudek:2016cru,Briceno:2016mjc}. To draw rigorous conclusions and determine whether there are bound states or resonances present, a L\"{u}scher analysis, where the finite-volume spectra are related to the scattering amplitudes, is necessary. To reliably constrain the scattering amplitudes, this would require calculations at non-zero momentum and/or different volumes which is beyond the scope of this first study. An important conclusion is that the addition of a class of operators resembling compact tetraquarks has little consequence on the finite-volume spectrum and, in turn, the scattering amplitudes.

\subsection{Doubly-charmed sector}
\label{subsec:results:doubly}

\begin{figure}[h]
\begin{center}
\includegraphics[width=0.9\textwidth]{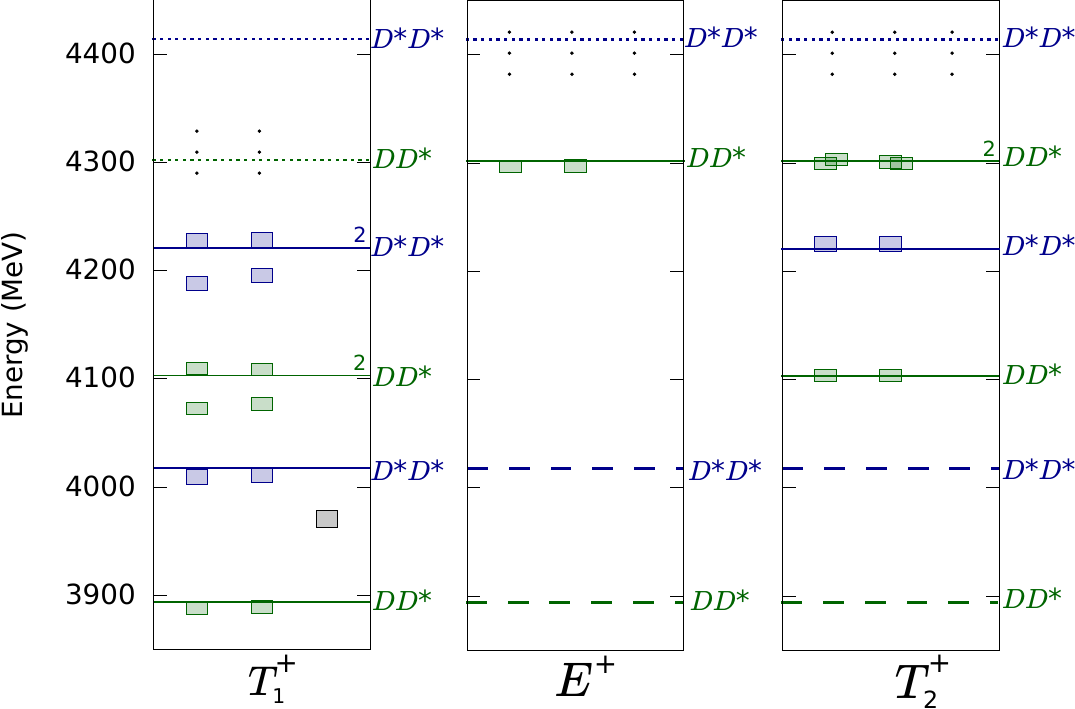}
\caption{As Figure~\ref{fig:hiddencharm} but for the isospin-0 doubly-charmed sector with quark flavour $cc\bar{l}\bar{l}$. Dashed lines indicate kinematic thresholds where a non-interacting level is not expected. Dotted lines indicate non-interacting meson-meson levels where the corresponding operators have not been included in the operator basis. Ellipses indicate that additional energy levels have been extracted in/above these regions but we have not plotted them as we have not included all relevant meson-meson operators in these energy regions.}
\label{fig:doublycharmed1}
\end{center}
\end{figure}

\begin{figure}[h]
\begin{center}
\includegraphics[width=0.7\textwidth]{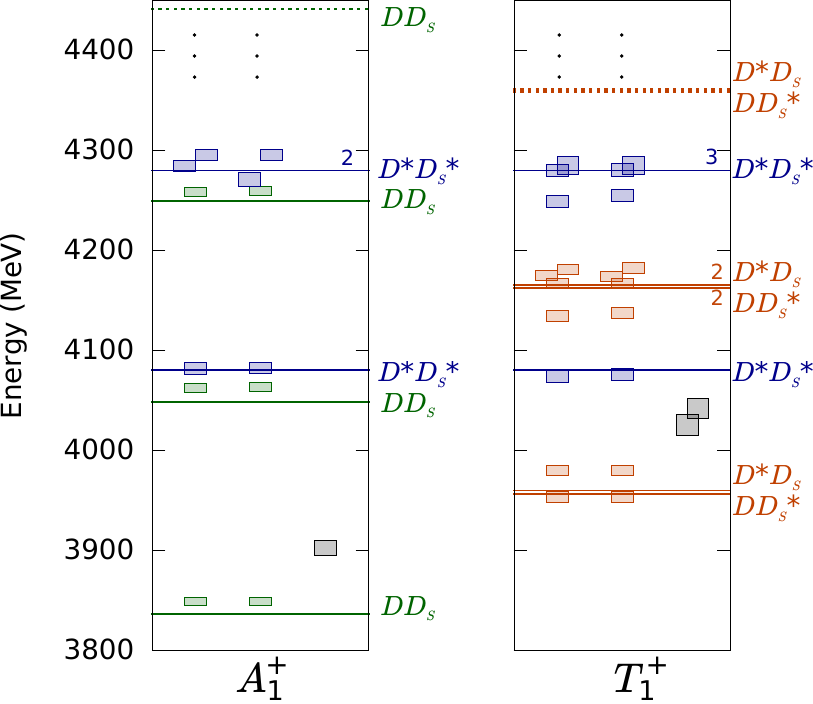}
\caption{As Figure~\ref{fig:doublycharmed1} but for the isospin-$\frac{1}{2}$ doubly-charmed sector with quark flavour $cc\bar{l}\bar{s}$.}
\label{fig:doublycharmed2}
\end{center}
\end{figure}

Turning to the doubly-charmed sector, Figure~\ref{fig:doublycharmed1} shows spectra for flavour content $cc\bar{l}\bar{l}$ in the $\Lambda^P = T_1^+, E^+, T_2^+$ isospin-0 channels\footnote{The lowest spin $J^P = 1^+$ appears in $T_1^+$ and the lowest spin $J^P = 2^+$ appears in $E^+, T_2^+$.} and Figure~\ref{fig:doublycharmed2} shows spectra for flavour $cc \bar{l}\bar{s}$ with isospin-$\frac{1}{2}$ in the irreps $\Lambda^P = A_1^+, T_1^+$. It can be again seen that there are no significant deviations between the spectra including and excluding tetraquark operators, and the spectra can not be reliably extracted using only tetraquark operators. Using the full basis of operators (see Table \ref{tab:doublecharmops}), the number of energy levels in each spectrum is equal to the number of expected non-interacting meson-meson energy levels in the relevant energy region. Because the basis of operators used has sufficiently diverse structures, we are able to extract many nearly-degenerate energy levels. In addition, we find that every energy level has a dominant meson-meson operator overlap. As in the results of the hidden-charm sector, we emphasise that the addition of a class of operators resembling compact tetraquarks does not significantly alter the finite-volume spectrum extracted.

The lowest-lying $DD$ and $D^\ast D^\ast$ levels in $s$-wave are forbidden in the $J^P=0^+,2^+$ isospin-0 doubly-charmed channels: the flavour wavefunction is antisymmetric in isospin-0 whilst the spin and spatial wavefunctions are symmetric, giving an overall antisymmetric wavefunction which is forbidden by Bose symmetry. These channels are particularly appealing to look for a tetraquark because if a low-lying state exists, it would lie far below the allowed non-interacting meson-meson energy levels and would be easily identified. Additionally, a low-lying $J^P=2^+$ stable tetraquark would subduce into both of the irreps $\Lambda^P = E^+, T_2^+$ and so appear with little ambiguity -- no such energy levels are seen in Figure~\ref{fig:doublycharmed1}. A $J^P=0^+$ tetraquark would appear in the $\Lambda^P = A_1^+$ irrep and, although a plot is not shown, we calculated the spectrum in this channel with the operators given in Table~\ref{tab:doublecharmops}. The first allowed non-interacting meson-meson energy level is $D \, D(2S)$, where $D(2S)$ is the first radial excitation of the $D$ meson.\footnote{We use a single-meson operator with structure $[\gamma^5 \overleftrightarrow{D} \overleftrightarrow{D}]$, where the two derivatives are coupled to $J=0$, for the $D(2S)$ rather than an optimised operator.} We do not find any energy levels below the $DD(2S)$ threshold at $\sim 4500$ MeV.

We now draw particular attention to the spectrum in the $\Lambda^P = T_1^+$ isospin-0 channel where the non-interacting $DD^\ast$ and $D^\ast D^\ast$ levels can have degeneracy two. We have reliably extracted two energy levels (the third and fourth) that have dominant overlap onto the two relevant $DD^\ast$ operators and two energy levels (fifth and sixth) that have dominant overlap onto the two relevant $D^\ast D^\ast$ operators. It can be seen that each pair of energy levels is non-degenerate which suggests there is some interaction. In order to quantify this, a further analysis requiring computations on different volumes and overall non-zero momentum is needed to relate the finite-volume spectrum to the scattering amplitudes via the L\"uscher formalism. It is also important to stress that a reliable determination of the coupled $s$ and $d$-wave scattering amplitudes in this channel depends on our ability to robustly extract these multiple energy levels.

%% file: stability.tex
\section{Systematics and stability of the extracted spectra}
\label{sec:stability}

Before discussing the results further, we consider some systematic effects which may have an impact on them and present some tests of varying the operator basis and the number of distillation vectors.

\subsection{Systematic uncertainties}
As a first application of these tetraquark operator constructions, we have performed calculations on a relatively small lattice volume, with spatial extent $L \sim 2$ fm, and this may be too small to distinguish the spatial structures of the extended meson-meson and compact tetraquark. The tetraquark operator can be Fierz rearranged as a linear combination of meson-meson operators multiplied by a factor of $1/L^{3}$~\cite{Padmanath:2015era}, suppressing the overlap of a possible tetraquark state onto the meson-meson operators. Further calculations, beyond the scope of this study, would be required to give some indication on how the results vary with the volume.

As this is a first demonstration, we have performed calculations with unphysically-heavy light quarks, corresponding to $m_\pi = 391$ MeV, and the presence/absence of tetraquarks may depend on the mass of the light quarks. Ultimately, calculations with light-quark masses approaching their physical values are required for comparison with experiment. On the other hand, studying how the spectra change as the quark masses are varied would give insight into the relevant QCD interactions and could be compared with expectations in different models.

Other possible systematic uncertainties include discretisation effects and the tuning of the charm quark mass. These issues were addressed in Ref.~\cite{Liu:2012} and we do not repeat the discussion here. 

\subsection{Varying the operator basis}

Some tests of how varying the operator bases affects the results have already been presented in Section~\ref{sec:results}. In summary, it was found that there were no significant changes in the low-lying spectra when only meson-meson operators were used compared to using the full basis of tetraquark and meson-meson operators. However, reliable spectra could not be extracted if only tetraquark operators were used.

\begin{figure}[h]
\begin{center}
\includegraphics[width=0.6\textwidth]{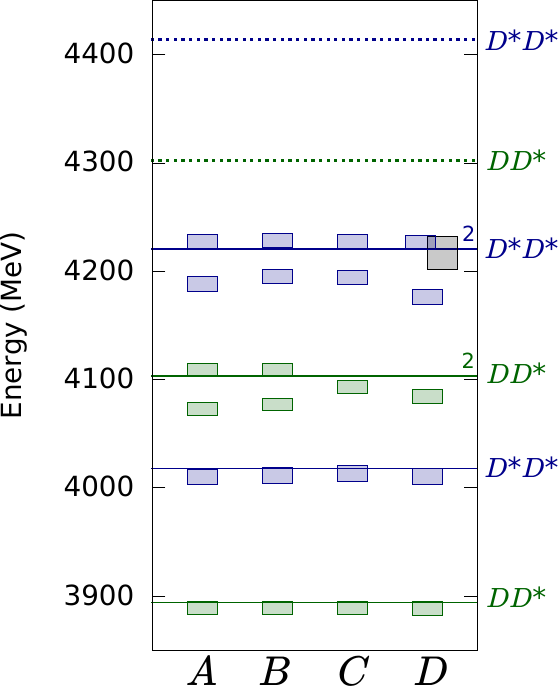}
\caption{As Figure~\ref{fig:doublycharmed1} but for the  $\Lambda^{P} = T_1^+$ isospin-0 $cc\bar{l}\bar{l}$ channel with different bases of operators: $A$ uses the full basis of meson-meson and tetraquark operators, $B$ uses only meson-meson operators, $C$ uses only meson-meson operators minus one $DD^\ast$ operator as described in the text, and $D$ uses the operators as in $C$ supplemented with the tetraquark operators.}
\label{fig:lessmes}
\end{center}
\end{figure}

As an illustration of what could happen if a sufficiently diverse set of meson-meson operators is not used, we show in Figure~\ref{fig:lessmes} spectra in the doubly-charmed isospin-0 $\Lambda^P = T_1^+$ channel computed using different operator bases. Note that degenerate meson-meson energy levels would be present here in the non-interacting limit as discussed in Section~\ref{subsec:mesonmeson}. Column $A$ shows the spectrum computed using the full basis of meson-meson and tetraquark operators (see Table~\ref{tab:doublecharmops}) and we see that the number of energy levels is equal to the number of expected non-interacting meson-meson energy levels in the energy region considered. We make the same conclusion for column $B$ which shows the spectrum calculated using only meson-meson operators. Column $C$ shows the spectrum using only meson-meson operators without the $D^{[100]}_{A_2} D^\ast{}_{E_2}^{[100]}$ operator which is relevant for the $DD^\ast$ level at $\approx 4100$ MeV -- it is seen that now one fewer energy level is extracted and the second $D D^\ast$ level moves slightly higher in energy which is as expected when not enough operators are used~\cite{Dudek:2012xn}. The right column $D$ shows the spectrum calculated with the operators as in $C$ supplemented with the tetraquark operators -- an additional level is found compared to $C$ high up in the spectrum. This demonstrates the necessity of accounting for all the relevant meson-meson energy levels in the energy region being considered and using a sufficient basis of operators of different structures. Otherwise there is the danger that this level could be mistakenly taken as a signal for the presence of a tetraquark.

\subsection{Varying the number of distillation eigenvectors}
\label{subsec:varydistillation}

\begin{figure}[h]
\begin{center}
\includegraphics[width=0.8\textwidth]{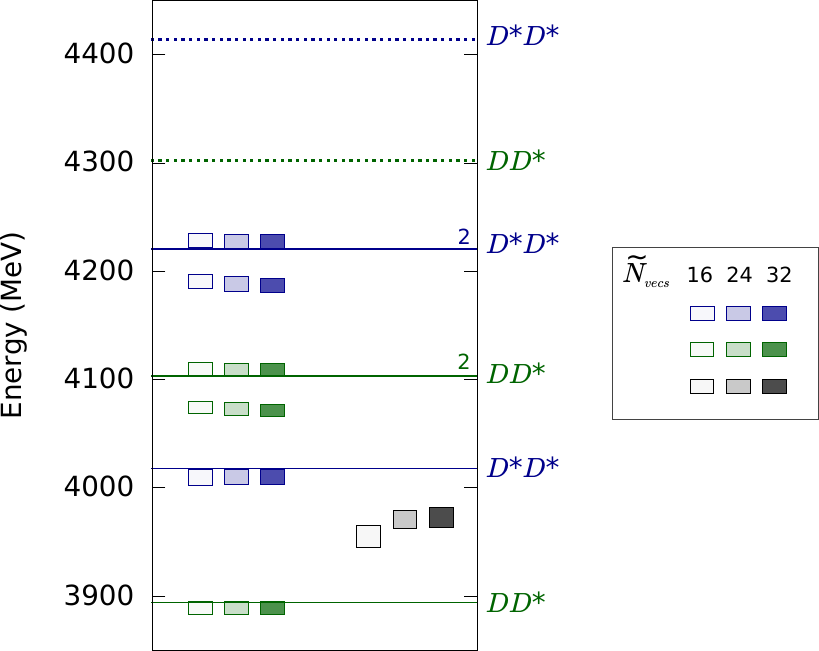}
\caption{As Figure~\ref{fig:doublycharmed1} but showing the $\Lambda^{P} = T_1^+$ isospin-0 $cc\bar{l}\bar{l}$ spectrum calculated using tetraquark operators with different numbers of distillation vectors, $\tilde{N}_{\text{vecs}} = 16,24,32$.  The left three columns are using the full basis of meson-meson and tetraquark operators, and the right three columns are using only tetraquark operators.}
\label{fig:hugedata}
\end{center}
\end{figure}

If the number of distillation eigenvectors used for the tetraquark operators is too low, the operator may not efficiently interpolate states of interest as it may be too smeared and so no longer resemble a compact tetraquark. However, as discussed in Section~\ref{subsec:distillation}, the computational cost involving tetraquark operators scales much more strongly than meson-meson operators with the number of distillation eigenvectors and therefore the number used can not be too large if the calculations are to be feasible. In this section, we test how sensitive the results are to varying number of distillation eigenvectors.

The spectrum in the doubly-charmed isospin-0 $\Lambda^{P} = T_1^+$ channel is shown in Figure~\ref{fig:hugedata} using different numbers of distillation vectors for tetraquark operators, $\tilde{N}_{\text{vecs}} = 16,24,32$, with both the full basis of meson-meson and tetraquark operators (see Table~\ref{tab:doublecharmops}) and only tetraquark operators. It can be seen that the results are not sensitive to the number of distillation eigenvectors used.

We also computed all the spectra using $N_{\text{vecs}} = \tilde{N}_{\text{vecs}} = 24$, i.e.\ the same number of distillation vectors for both meson-meson and tetraquark operators. The results were found to be consistent with the spectra presented in Section~\ref{sec:results} (which used $\tilde{N}_{\text{vecs}} = 24$ for tetraquark operators and $N_{\text{vecs}} = 64$ for meson-meson operators). This shows that the results are also not sensitive to the number of distillation vectors used for the meson-meson operators.

In summary, these tests suggest that the results are not very sensitive to the number of distillation vectors being used. In addition, a recent study in Ref.~\cite{Woss:2016tys} demonstrated that a small number of distillation vectors is sufficient to extract finite-volume spectra as long as one does not consider higher momenta, higher spin or highly excited states. Because we are considering overall zero momentum and relatively low-lying states, this gives further support to our conclusion.

%% file: interpretations.tex
\section{Discussion and comparison with previous studies}
\label{sec:interpretations}

In this section we discuss the results in the context of expectations from phenomenological models and compare with previous lattice calculations. In a simple one-gluon-exchange model of a diquark as described in Appendix~\ref{subsec:onegluon}, the two quarks interact via a colour-colour spin-spin interaction and the most attractive diquark and anti-diquark configurations have (colour irrep, spin) $= (\underline{\bar{3}},0)$ and $(\underline{3},0)$ respectively. Therefore, the most favourable tetraquark has $J^{P} = 0^{+}$ which subduces into $\Lambda^{P} = A_1^{+}$. However, a large quark mass suppresses the spin-spin interactions and so some models expect spin-1 diquark configurations involving heavy quarks to occur and form a tetraquark multiplet~\cite{Maiani:2004vq}. This multiplet contains tetraquarks with $J^P=1^+$ and $2^+$ which subduce into the $T_1^+$ irrep and the $E^+,T_2^+$ irreps respectively. Besides the quark mass, this one-gluon-exchange interaction does not depend on the flavours of the quarks.  However, when the flavour irreps of the two quarks (antiquarks) are the same, Fermi symmetry requires the overall diquark (anti-diquark) configurations to be antisymmetric and this restricts the allowed structures.

In the hidden-charm isospin-1 sector, there are no identical quarks/antiquarks and so no constraints from symmetry on the allowed configurations. Models~\cite{Maiani:2004vq,Maiani:2014aja} suggest that the lightest tetraquark multiplet has $J^{PG} = 0^{+-}, 1^{++}, 1^{+-}, 2^{+-}$ and we have performed a thorough investigation in all these channels except for $J^{PG}=2^{+-}$. In the $\Lambda^{PG} = A_1^{+-}$ channel, expected to be the most attractive, and the $\Lambda^{PG} = T_1^{++}$ and $T_1^{+-}$ channels, there are no hints of a narrow state or any significant interactions in the computed spectra. No experimental candidate has been observed with $J^{PG}=0^{+-}$, nor is there currently any charged charmonium-like candidate with undetermined $J^{PG}$ that is light enough to be identified as the lowest-lying $J^{PG} = 0^{+-}$ tetraquark. The observed $Z_c^+(3900)$ has $J^{PG} = 1^{++}$~\cite{Olive:2016xmw} and has been suggested to be a candidate for a tetraquark. That we see no sign of it is consistent with previous lattice QCD calculations presented in Ref.~\cite{Prelovsek:2014swa} which also calculated the finite volume spectrum using meson-meson and tetraquark operators. Our results are also consistent with other lattice QCD calculations \cite{Chen:2014afa,Chen:2015jwa,Ikeda:2016zwx} which do not find evidence of a bound state or narrow resonance in this channel. There is currently no well-established experimental candidate with $J^{PG} = 1^{+-}$, but if the $X(3872)$ is a tetraquark its isospin-1 partner would appear in this channel. Again, that we see no clear signal for a state here is consistent with previous lattice QCD calculations~\cite{Padmanath:2015era}. That study also found that the spectrum in this channel was insensitive to the addition of tetraquark-like operators to the operator basis. 

In the doubly-charmed sector, possible diquark configurations are further constrained by Fermi statistics. The $(\bar{3},0)$ and $(6,1)$ $cc$ diquarks are forbidden as they are symmetric under the interchange of quarks and only the $(\bar{3},1)$ and $(6,0)$ diquarks are allowed. In the one-gluon exchange model given in Appendix~\ref{subsec:onegluon}, the colour-colour spin-spin interaction is repulsive for these allowed configurations and is least repulsive for $( \bar{\underline{3}},1)$. The attractive $(\underline{3},0)$ $\bar{q}\bar{q}$ anti-diquark configurations are required to be antisymmetric in flavour in $F=\underline{3}$ and the most attractive configuration has isospin, $I=0$. Therefore, the most favourable tetraquark has $(I)J^P = (0)1^+$. Other attractive configurations include $(I)J^P = (0)0^+$, $(0)2^+$ containing a $(\underline{\bar{6}},1)$ anti-diquark and $(I)J^P = (\tfrac{1}{2})0^+, (\tfrac{1}{2})1^+$ from picking the $I=\tfrac{1}{2}$ components of the $(\underline{3},0)$ anti-diquark. However, no signs of these are seen in any of the computed spectra in the many doubly-charmed channels we studied. That we find no significant deviation between the spectra including and excluding tetraquark operators is consistent with the results presented in Ref.~\cite{Guerrieri:2014nxa} which computed the spectrum in the $(I)J^P = (0)1^+$ channel. That study used meson-meson and tetraquark operators but, because the operator basis was more restricted than ours, was unable to extract all of the multiple levels which correspond to degenerate meson-meson levels in the non-interacting limit. Computations presented in Ref.~\cite{Ikeda:2013vwa} find an attractive interaction in the $(I)J^P = (0)1^+$ channel using a less direct approach in which lattice QCD computations are used to extract a potential which is then used to determine scattering amplitudes. They do not find a bound state or resonance for a range of light quark masses corresponding to $m_\pi = 410 - 700$ MeV and conclude that this attractive interaction gets stronger with decreasing pion mass, further motivating studies of how the results vary as the light quark mass decreases towards the physical point.

In one-gluon exchange models, the colour-colour spin-spin interaction is always repulsive for the $cc$ diquark, but the repulsion is suppressed by the quark mass which suggests that doubly-bottomed tetraquarks may be more favourable than doubly-charmed tetraquarks. This is supported by lattice QCD calculations of finite volume-spectra using bases of meson-meson and tetraquark-like operators which suggest the existence of a $(I)J^P = (0)1^+$ doubly-bottomed tetraquark~\cite{Francis:2016hui}. Further support comes from lattice calculations of the potential between two static bottom quarks in the presence of two light antiquarks~\cite{Bicudo:2012qt,Bicudo:2015vta, Bicudo:2015kna, Peters:2016isf, Bicudo:2017szl}. This potential is found to lead to a bound state with $(I)J^P = (0)1^+$. Our doubly-charmed $(I)\Lambda^P = (0)T_1^+$ spectrum is not inconsistent with there being an attractive interaction although there were no obvious signs of a bound state in this channel. This is also consistent with recent phenomological studies \cite{Karliner:2017qjm,Eichten:2017ffp,Czarnecki:2017vco} which suggest the doubly-bottom tetraquark is bound and the doubly-charmed tetraquark is unbound. Further calculations using bottom quarks and a L\"{u}scher analysis would be of interest. Computations involving the bottom quark with the fermion action used in this study are not straightforward since discretisation effects would be large. It is possible to implement the bottom quark with alternative actions such as Non-Relativistic QCD but this is beyond the scope of this study.

Overall, our study has improved on previous lattice QCD investigations of tetraquarks in two ways. The first is that we use a diverse set of tetraquark and meson-meson operators so that we can reliably obtain a large number of energy levels in each channel and, for the first time in a lattice QCD calculation, robustly extract the multiple energy levels associated with meson-meson energy levels which are degenerate in the non-interacting limit.\footnote{Recall that, as discussed in Section~\ref{subsec:mesonmeson}, these can occur when at least one of the mesons has non-zero spin.} This is important for future spectrum calculations involving the scattering of mesons with non-zero spin. The second is that we have computed spectra in a large number of channels proposed to contain the hypothetical lightest tetraquark multiplet -- some of these channels have not been studied before.

%% file: conclusions.tex
\section{Summary}\label{sec:summary}

We have described the construction of a general class of operators resembling compact tetraquarks which have a range of different diquark--anti-diquark structures, transform irreducibly under the symmetries of the lattice and respect other relevant symmetries. As a first demonstration, these operators have been used in conjunction with meson-meson operators to compute correlation functions in the isospin-1 hidden-charm and doubly-charmed sectors using the distillation framework. Finite-volume spectra were extracted by analysing the correlation functions with the variational method. It was found that the addition of tetraquark operators to a basis of meson-meson operators did not significantly affect the finite-volume spectrum and subsequently, would not affect the scattering amplitudes. Because a diverse set of operators was used, for the first time we were able to reliably extract the multiple energy levels associated with degenerate non-interacting meson-meson levels. In all channels, we find that the number of energy levels is equal to the number of non-interacting meson-meson levels expected in the energy region considered and the majority of energies were at most slightly shifted from the non-interacting levels. Hence, there are no strong indications that there are any bound states or narrow resonances present. 

This study sets out the groundwork and technology for future work. Calculations with larger lattice volumes and/or at non-zero overall momentum would be necessary to reliably determine scattering amplitudes via the L\"{u}scher method and so rigorously discern the bound state and resonance content in the various channels. In addition, there is strong motivation to study how the results change when the light quark mass is varied. Calculations with lower quark masses would require large volumes. As discussed in Ref.~\cite{Peardon:2009}, to maintain a given smearing radius in the distillation approach, the number of distillation vectors used must scale with the volume. Because tetraquark elementals are of rank-4, increasing the number of distillation vectors vastly increases the computational cost and so, to make the calculations feasible, an extension of the distillation framework would be required, for example, a stochastic version~\cite{Morningstar:2011ka} or an alternative basis of vectors. Calculations with more distillation vectors would also enable the dependence of the results on the degree of tetraquark-operator smearing to be investigated further. Whilst the extracted spectra in the channels studied did not significantly change upon the addition of tetraquark operators to the operator basis, there are many other channels where tetraquarks have been suggested to exist and more detailed lattice QCD investigations are of interest, for example, in the isospin-0 hidden-charm sector, the open-charm sector, the bottom sector and the light scalar mesons.

%% file: appendix.tex
\section*{Appendices}
\section{Diquark and tetraquark operators} \label{app:tetraquark}
In this appendix we present some additional details of the diquark and tetraquark operators. Consider a diquark operator, $\delta(\vec{x},t) = \sum \text{CGs} \ q^T(\vec{x},t) (C \Gamma) q (\vec{x},t)$, where CGs refers to the Clebsch-Gordan coefficients in Equation~\eqref{eqn:diquark} and various indices have been suppressed. Under proper Lorentz transformations, this operator transforms in the same way as the analogous fermion bilinear and the continuum spin, $J$, of the diquark for different choices of $\Gamma$ is given in Table~\ref{tab:gamma}. Under a parity transformation, the operator transforms to $\mathcal{P}\delta(\vec{x},t)\mathcal{P}^{-1} = \eta_P q^T(-\vec{x},t) C \Gamma q(-\vec{x},t)$, where $\eta_P = \pm 1 $ is a parity factor that depends on the gamma matrix $\Gamma$ as given in the table. The analogous anti-diquark operator has the same transformation properties. The parity of the tetraquark operator is the product of the parity factors of the diquark and anti-diquark operators.

Taking the Hermitian conjugate of the diquark, we obtain $\delta^\dagger = h_\Gamma s_C s_F \bar{\delta}$, where the symmetry of the Dirac gamma matrix $h_\Gamma = \pm 1$ is shown in Table~\ref{tab:gamma} and $s = \xi_1 \xi_3$ are phases arising from the exchange symmetry of the colour $(s_C)$ and flavour $(s_F)$ Clebsch-Gordan coefficients of the diquark operator. The phase $\xi_1 = \pm 1$ arises from reversing the order of the SU(3) irreps,
 \begin{equation}\langle D_1, d_1; D_2, d_2 | D, d \rangle = \xi_1 \langle D_2, d_2; D_1, d_1 | D, d \rangle, \end{equation}
 and the phase $\xi_3 = \pm 1$ arises from complex conjugating the irreps, 
\begin{equation}\langle D_1, d_1; D_2, d_2 | D, d \rangle = \xi_3 \langle \bar{D}_1, \bar{d}_1; \bar{D}_2, \bar{d}_2 | \bar{D}, \bar{d} \rangle.\end{equation}
  We use the phase conventions of Refs.~\cite{deSwart:1963pdg,Kaeding:1995vq}. For tetraquark operators with $G$-parity symmetry as in Equation~\eqref{eqn:tetragparity}, the $G$-parity is given by $G = \tilde{G} \xi_J \xi_1 \xi_3$ where $\xi_J = (-1)^{J_1 + J_2 - J}$ is the phase arising when the arguments of the $SU(2)$ Clebsch-Gordan coefficients are interchanged and $\xi_1$ and $\xi_3$ here arise from the exchange symmetry of the $SU(3)_F$ Clebsch-Gordan coefficients of the tetraquark operator.

When the flavour irreps of the quarks in the diquark are identical, the overall colour-flavour-spin coupling in the diquark must be antisymmetric due to Fermi statistics.  To see this schematically, consider the diquark $\delta =  \sum C_{ab} \ q_a q_b$ with overall coupling coefficients $C$. If $C_{ab}$ is symmetric, the sum would be exactly zero because $q_a q_b$ is antisymmetric. The symmetry arising from spin $(C\Gamma)_{\alpha \beta} = s_\Gamma (C\Gamma)_{\beta \alpha}$ is given in Table~\ref{tab:gamma} and the symmetries arising from colour and flavour are discussed above.

\begin{table}[h!]
\begin{center}
\begin{tabular}{c|cccccccc}
 & 1 & $\gamma_5$ & $\gamma_0 \gamma_5$ & $\gamma_0$ & $\gamma_i$ & $\gamma_i \gamma_0$ & $\gamma_5 \gamma_i$ & $[\gamma_i,\gamma_j]$\\ 
\hline
$\Gamma$ & $a_0$ & $\pi$ & $\pi_2$ & $b_0$ & $\rho$ & $\rho_2$ & $a_1$ & $b_1$ \\
\hline
$J$ & 0 & 0 & 0 & 0 & 1 & 1 & 1 & 1 \\
\hline
$\eta_P$ & $-$ & $+$ & $+$ & $-$ & $+$ & $+$ & $-$ & $-$ \\
\hline
$h_\Gamma$ & $+$ & $-$ & $+$ & $+$ & $+$ & $-$ & $+$ & $-$\\
\hline
$s_\Gamma$ & $-$ & $-$ & $-$ & $+$ & $+$ & $+$ & $-$ & $+$
\end{tabular}
\caption{For different Dirac gamma matrices, we show the notation $\Gamma$ used to denote the gamma matrix, the continuum spin $J$, the parity factor $\eta_P$, the hermiticity factor $h_\Gamma$ and the spin coupling symmetry $s_\Gamma$.}
\label{tab:gamma}
\end{center}
\end{table}

\section{One-gluon exchange model}
\label{subsec:onegluon}

In a simple one-gluon exchange model of a diquark \cite{Jaffe:2004ph}, the two quarks interact via a colour-colour spin-spin interaction term,
\begin{equation}
\label{equ:onegluon}
  H =  - \alpha_s A_{12}(\lambda_1 \cdot \lambda_2) (\vec{S}_1 \cdot \vec{S}_2)
\end{equation}
where $A_{12}$ is a model-dependent mass term that behaves like $1/m_1 m_2$ in the heavy quark limit, $\lambda$ are the Gell-Mann matrices that span the Lie algebra of SU$(3)_C$ and $\vec{S}$ is the spin of the quark. The relative factors that arise for various colour irreps $R$ and spin $S$ are given in Table~\ref{tab:onegluon}. It can be seen that the most attractive diquark is the $(R,S) =  (\underline{\bar{3}}, 0)$ configuration. Similarly, the most attractive anti-diquark is the $(\underline{3},0)$ configuration. Hence a scalar $J^P = 0^+$ tetraquark is expected to be the most favourable. Whilst other configurations are less favourable, this one-gluon exchange interaction is suppressed by the masses of the quarks such that in the heavy quark limit, a rich spectrum of tetraquark states with $J^P = 0^+, 1^+, 2^+$ is expected to be observed in models such as Ref.~\cite{Maiani:2004vq}.

In the case when the flavour irreps of the quarks within the diquark are identical, Fermi symmetry constrains the number of possible configurations. If the flavour irrep is antisymmetric, then the only allowed diquarks are the attractive configurations, $(\underline{\bar{3}}, 0)$ and $(\underline{6}, 1)$. On the other hand, when the flavour irrep is symmetric the only allowed diquarks are the repulsive configurations, $(\underline{\bar{3}}, 1)$ and $(\underline{6}, 0)$. A consequence of this is that the doubly-charmed $cc$ diquark is always repulsive with the least repulsive diquark being $(\underline{\bar{3}},1)$. However, the repulsive interaction is suppressed by the quark mass and so it is expected that such tetraquarks may exist in the heavy quark limit~\cite{MANOHAR199317}.

\begin{table}[h!]
\begin{center}
\begin{tabular}{cc|cc}
&& \multicolumn{2}{c}{$S$}\\
& & 0 & 1 \\ \hline 
\multirow{2}{*}{R}& $\underline{\bar{3}}$ & $\frac{1}{2}$ & $-\frac{1}{6}$ \\[0.25cm]
& $\underline{6}$ & $-\frac{1}{4}$ & $\frac{1}{12}$
\end{tabular}
\caption{The relative factors of the colour-colour spin-spin interaction within the diquark in equation~\eqref{equ:onegluon} for various color irreps and spin. }
\label{tab:onegluon}
\end{center}
\end{table}

\section{Non-relativistic quark model}
\label{subsec:nonrelativistic}

In a non-relativistic quark model, diquark states at rest with orbital angular momentum $L$ and spin angular momentum $S$ coupled to total angular momentum $J$ can be constructed as,
\begin{equation}
\big|\delta^{J,m}_{LS} \big\rangle = \sum_{m_L,m_S} \langle L, m_L; S, m_S| J,m \rangle \sum_{\alpha, \beta} \big\langle \tfrac{1}{2}, \alpha; \tfrac{1}{2}, \beta \big| S, m_S \big\rangle 
\int\! \frac{d^3 q}{(2\pi)^3} \ Y^{m_L}_L (\hat{q}) f_{nL}(|\vec{q}|) b^\dagger_\alpha(\vec{q}) b^\dagger_\beta(-\vec{q}) | 0 \rangle \, ,
\end{equation}
where $b_\alpha^\dagger(\vec{q})$ is a creation operator for a quark of momentum $\vec{q}$ and $J_z$ component $\alpha$, and $f_{nL}(|\vec{q}|)$ is a model-dependent wavefunction that is determined by some interaction potential and is specified by $L$ and the principal quantum number $n$. Annihilating this state with the field expansion of the diquark operator, we obtain,
\begin{equation}
\begin{aligned}
  \big\langle 0 \big| \delta^{J[\Gamma]} \big|\delta^{J,m}_{LS} \big\rangle =& \sum_{m_L,m_S} \langle L, m_L; S,m_S| J,m \rangle \sum_{\alpha,\beta} \big\langle \tfrac{1}{2}, \alpha; \tfrac{1}{2}, \beta \big| S, m_S \big\rangle \\
  & \times \int \! \frac{d^3 q}{(2\pi)^3} \ Y^{m_L}_L (\hat{q}) f_{nL}(|\vec{q}|) \, u^T_{(\alpha)}(\vec{q}) C\Gamma u_{(\beta)} (-\vec{q}) \, ,
\end{aligned}
\end{equation}
where $u$ is a Dirac spinor. Expanding $u$ in the non-relativistic limit where $|\vec{q}|$ is much smaller than the mass of the quark, we find to leading order for $\Gamma = \gamma^5$, 
\begin{equation} 
\begin{aligned}
\big\langle 0 \big| \delta^{J[\gamma^5]} \big|\delta^{J,m}_{LS} \big\rangle =& \sum_{m_L,m_S} \langle L, m_L; S,m_S| J,m \rangle \\ &\times \underbrace{\left( \big\langle \tfrac{1}{2}, -\tfrac{1}{2}; \tfrac{1}{2}, \tfrac{1}{2} \big| S, m_S \big\rangle - \big\langle \tfrac{1}{2}, \tfrac{1}{2}; \tfrac{1}{2}, -\tfrac{1}{2} \big| S, m_S \big\rangle \right)}_{\sim\delta_{S0} \delta_{m_S 0}}  \underbrace{\int \! \frac{d^3 q}{(2\pi)^3} \ Y^{m_L}_L (\hat{q})}_{\sim \delta_{L0}\delta_{m_L0}} f_{nL}(|\vec{q}|) \, .
\end{aligned}
 \end{equation}
Hence, $\delta^{J[\gamma^5]}$ overlaps with the $qq({}^{2S+1}\!L_J={}^1\!S_0)$ diquark construction. Similar results for other $\Gamma$ are shown in Table~\ref{tab:nonrel}.

\begin{table}[h!]
\begin{center}
\begin{tabular}{c|cccccccc}
 & 1 & $\gamma_5$ & $\gamma_0 \gamma_5$ & $\gamma_0$ & $\gamma_i$ & $\gamma_i \gamma_0$ & $\gamma_5 \gamma_i$ & $[\gamma_i,\gamma_j]$\\ 
\hline
 $qq({}^{2S+1}\!L_J)$ & ${}^3\!P_0$ & ${}^1\!S_0$ & ${}^1\!S_0$ & - & ${}^3\!S_1$ & ${}^3\!S_1$ & ${}^3\!P_1$ & ${}^1\!P_1$ \\
\end{tabular}
\caption{The non-relativistic overlap of the diquark operator $\delta^{J[\Gamma]}$ onto the diquark state $qq({}^{2S+1}\!L_J)$.}
\label{tab:nonrel}
\end{center}
\end{table}

\section{Operator lists}\label{app:ops}
The interpolating operators used to calculate the spectra are listed in Table~\ref{tab:hiddencharmops} for the hidden-charm sector and Table~\ref{tab:doublecharmops} for the doubly-charmed sector. 

\begin{table}[h!]
\begin{center}
\begin{tabular}{c|c|c}
\hline
 $T_1^{++}$ & $A_1^{+-}$ &$T_1^{+-}$\\
\hline
$\delta^{b_0}_{\bar{3},3} \bar{\delta}^{b_1}_{3,\bar{3}}$  & $\delta^{a_0}_{\bar{3},3} \bar{\delta}^{a_0}_{3,\bar{3}}$ & $\delta^{a_0}_{\bar{3},3} \bar{\delta}^{a_1}_{3,\bar{3}}$  \\
$\delta^{\rho}_{\bar{3},3} \bar{\delta}^{\pi}_{3,\bar{3}}$  & $\delta^{\pi}_{\bar{3},3} \bar{\delta}^{\pi}_{3,\bar{3}}$ & $\delta^{\rho}_{\bar{3},3} \bar{\delta}^{\pi}_{3,\bar{3}}$  \\
$\delta^{\rho}_{\bar{3},3} \bar{\delta}^{\rho}_{3,\bar{3}}$  & $\delta^{\rho}_{6,3} \bar{\delta}^{\rho}_{\bar{6},\bar{3}}$ & $\delta^{\rho}_{\bar{3},3} \bar{\delta}^{\rho_2}_{3,\bar{3}}$  \\
$D^{[000]}_{A_1} \bar{D}^\ast{}^{[000]}_{T_1}$ &  $D^{[000]}_{A_1} \bar{D}^{[000]}_{A_1}$ & $D^{[000]}_{A_1} \bar{D}^\ast{}^{[000]}_{T_1}$ \\
  $D^\ast{}^{[000]}_{T_1} \bar{D}^\ast{}^{[000]}_{T_1}$  & $D^{[100]}_{A_2} \bar{D}^{[100]}_{A_2}$ & $D^{[100]}_{A_2} \bar{D}^\ast{}^{[100]}_{A_1}$ \\
$\eta_c{}^{[000]}_{A_1} \rho^{[000]}_{T_1}$ & $D^\ast{}^{[000]}_{T_1} \bar{D}^\ast{}^{[000]}_{T_1}$ & $D^{[100]}_{A_2} \bar{D}^\ast{}^{[100]}_{E_2}$ \\
$J/\psi^{[000]}_{T_1} \pi^{[000]}_{A_1}$ & $\eta_c{}^{[000]}_{A_1} \pi^{[000]}_{A_1}$ & $J/\psi^{[000]}_{T_1} \rho^{[000]}_{T_1}$ \\
$J/\psi^{[100]}_{A_1} \pi^{[100]}_{A_2}$ & $\eta_c{}^{[100]}_{A_2} \pi^{[100]}_{A_2}$ & $J/\psi^{[100]}_{A_1} \rho^{[100]}_{E_2}$ \\
$J/\psi^{[100]}_{E_2} \pi^{[100]}_{A_2}$ & $J/\psi^{[000]}_{T_1} \rho^{[000]}_{T_1}$ & $J/\psi^{[100]}_{E_2} \rho^{[100]}_{A_1}$ \\
 &  & $J/\psi^{[100]}_{E_2} \rho^{[100]}_{E_2}$\\
 && $\chi_{c0}{}^{[100]}_{A_1} \pi^{[100]}_{A_2}$
\end{tabular}
\caption{The interpolating operators used to calculate the spectra in the isospin-1 hidden-charm sector. For the tetraquark operators, we use the notation $\delta_{R_1,F_1}^{\Gamma_1} \bar{\delta}_{R_2,F_2}^{\Gamma_2}$ where $R_1 (R_2)$ is the colour irrep, $\Gamma_1 (\Gamma_2)$ is the gamma matrix and $F_1 (F_2)$ is the flavour irrep of the diquark (anti-diquark) operator. For meson-meson operators, the optimised single-meson operators used are denoted by $M_\Lambda^{[n_1n_2n_3]}$, where $M$ indicates the meson, $\Lambda$ is the lattice irrep and $[n_1n_2n_3]$ is the momentum in units of $\frac{2\pi}{L}$. Note that all momenta related to $[n_1n_2n_3]$ by an allowed lattice rotation are summed over as shown in Equation~\eqref{eqn:mesonmeson}.}
\label{tab:hiddencharmops}
\end{center}
\end{table}

\begin{table}[h!]
\begin{center}
\begin{tabular}{c|c|c|c || c |c}
\multicolumn{4}{c||}{$I=0$} & \multicolumn{2}{c}{$I=\frac{1}{2}$} \\
\hline
$A_1^{+}$ & $T_1^{+}$ & $E^{+}$ & $T_2^{+}$ & $A_1^+$ & $T_1^+$ \\
\hline
$\delta^{a_0}_{6,1} \bar{\delta}^{b_0}_{\bar{6},3}$ & $\delta^{b_1}_{\bar{3},1} \bar{\delta}^{a_0}_{3,3}$ &  $\delta^{a_1}_{6,1} \bar{\delta}^{b_1}_{\bar{6},3}$  & $\delta^{a_1}_{6,1} \bar{\delta}^{b_1}_{\bar{6},3}$ & $\delta^{b_0}_{\bar{3},1} \bar{\delta}^{b_0}_{3,\bar{6}}$  & $\delta^{a_1}_{6,1} \bar{\delta}^{b_0}_{\bar{6},3}$\\
$\delta^{a_1}_{6,1} \bar{\delta}^{b_1}_{\bar{6},3}$ & $\delta^{\rho_2}_{\bar{3},1} \bar{\delta}^{\pi_2}_{3,3}$ & $\delta^{b_1}_{\bar{3},1} \bar{\delta}^{a_1}_{3,3}$ & $\delta^{b_1}_{\bar{3},1} \bar{\delta}^{a_1}_{3,3}$ & $\delta^{b_1}_{\bar{3},1} \bar{\delta}^{a_1}_{3,3}$ & $\delta^{b_1}_{\bar{3},1} \bar{\delta}^{a_0}_{3,3}$\\
$\delta^{b_0}_{\bar{3},1} \bar{\delta}^{a_0}_{3,3}$ &$\delta^{\rho}_{\bar{3},1} \bar{\delta}^{\pi_2}_{3,3}$ & $D^{[110]}_{A_2} D^\ast{}^{[110]}_{B_2} $ & $D^{[100]}_{A_2} D^\ast{}^{[100]}_{E_2}$ &$\delta^{\pi}_{6,1} \bar{\delta}^{\pi}_{\bar{6},\bar{6}}$ & $\delta^{\rho}_{\bar{3},1} \bar{\delta}^{\pi}_{3,3}$\\
$\delta^{b_1}_{\bar{3},1} \bar{\delta}^{a_1}_{3,3}$ &$\delta^{\rho}_{\bar{3},1} \bar{\delta}^{\pi}_{3,3}$ & & $D^{[110]}_{A_2} D^\ast{}^{[110]}_{A_1}$ & $\delta^{\rho}_{\bar{3},1} \bar{\delta}^{\rho}_{3,\bar{6}}$ &$\delta^{\rho}_{\bar{3},1} \bar{\delta}^{\rho}_{3,\bar{6}}$ \\
$D_{A_1}^{[000]} D(2S)_{A_1}^{[000]}$ & $D^{[000]}_{A_1} D^\ast{}^{[000]}_{T_1}$ & & $D^{[110]}_{A_2} D^\ast{}^{[110]}_{B_1}$ & $D^{[000]}_{A_1} D_s{}^{[000]}_{A_1}$ & $D^{[000]}_{A_1} D_s^\ast{}^{[000]}_{T_1}$\\
& $D^{[100]}_{A_2} D^\ast{}^{[100]}_{A_1}$ & & $D^\ast{}^{[100]}_{A_1} D^\ast{}^{[100]}_{E_2}$ & $D^{[100]}_{A_2} D_s{}^{[100]}_{A_2}$ & $D^\ast{}^{[000]}_{T_1} D_s{}^{[000]}_{A_1}$\\
& $D^{[100]}_{A_2} D^\ast{}^{[100]}_{E_2}$ && & $D^{[110]}_{A_2} D_s{}^{[110]}_{A_2}$  & $D^{[100]}_{A_2} D_s^\ast{}^{[100]}_{A_1}$\\
&$D^\ast{}^{[000]}_{T_1} D^\ast{}^{[000]}_{T_1}$ & & & $D^\ast{}^{[000]}_{T_1} D_s^\ast{}^{[000]}_{T_1}$ & $D^{[100]}_{A_2} D_s^\ast{}^{[100]}_{E_2}$\\
& $D^\ast{}^{[100]}_{A_1} D^\ast{}^{[100]}_{E_2}$ & & & $D^\ast{}^{[100]}_{A_1} D_s^\ast{}^{[100]}_{A_1}$ & $D^\ast{}^{[100]}_{A_1} D_s{}^{[100]}_{A_2}$\\
&$D^\ast{}^{[100]}_{E_2} D^\ast{}^{[100]}_{E_2}$ & & & $D^\ast{}^{[100]}_{E_2} D_s^\ast{}^{[100]}_{E_2}$ & $D^\ast{}^{[100]}_{E_2} D_s{}^{[100]}_{A_2}$\\
&& & && $D^\ast{}^{[000]}_{T_1} D_s^\ast{}^{[000]}_{T_1}$\\
&& & && $D^\ast{}^{[100]}_{A_1} D_s^\ast{}^{[100]}_{E_2}$\\
&&&&& $D^\ast{}^{[100]}_{E_2} D_s^\ast{}^{[100]}_{A_1}$ \\
&&&&& $D^\ast{}^{[100]}_{E_2} D_s^\ast{}^{[100]}_{E_2}$ \\
\end{tabular}
\caption{As Table~\ref{tab:hiddencharmops} but for the doubly-charmed sector with isospin-0 (left columns) and isospin-$\frac{1}{2}$ (right columns).}
\label{tab:doublecharmops}
\end{center}
\end{table}